\documentclass[preprint,12pt,3p]{elsarticle}

\usepackage{amssymb}
\usepackage{float}

\journal{Methods in Enzymology}

\begin{document}

\begin{frontmatter}

\textit{Running title: Simulating Quantum Effects in Enzymes}
\title{Simulating nuclear and electronic quantum effects in enzymes}

\author[label1]{Lu Wang}
\address[label1]{Department of Chemistry and Chemical Biology, Rutgers University, Piscataway, New Jersey 08854, USA}
\ead{lwang@proteomics.rutgers.edu}

\author[label2]{Christine M. Isborn}
\address[label2]{Chemistry and Chemical Biology, School of Natural Sciences, University of California, Merced, 5200 North Lake Road, Merced, California 95343, USA}
\ead{cisborn@ucmerced.edu}

\author[label3]{Thomas E. Markland}
\address[label3]{Department of Chemistry, Stanford University, Stanford, California 94305, USA}
\ead{tmarkland@stanford.edu}

\begin{abstract}
An accurate treatment of the structures and dynamics that lead to enhanced chemical reactivity in enzymes requires explicit treatment of both electronic and nuclear quantum effects. The former can be captured in \textit{ab initio} molecular dynamics (AIMD) simulations while the latter can be included by performing \textit{ab initio} path integral molecular dynamics (AI-PIMD) simulations. Both AIMD and AI-PIMD simulations have traditionally been computationally prohibitive for large enzymatic systems. Recent developments in streaming computer architectures and new algorithms to accelerate path integral simulations now make these simulations practical for biological systems, allowing elucidation of enzymatic reactions in unprecedented detail. In this chapter, we summarize these recent developments and discuss practical considerations for applying AIMD and AI-PIMD simulations to enzymes.
\end{abstract}


\end{frontmatter}

\section{Introduction}
Tremendous effort has been devoted to using molecular simulation to unravel how enzymes catalyze chemical reactions with such remarkable efficiency and selectivity. A large amount of this work has utilized classical molecular mechanical (MM) methods combined with fixed charge and, more recently, polarizable empirical force fields \citep{Ponder2003}. These classical methods have been successful in predicting and elucidating important properties ranging from ligand binding free energies to $pK_a$'s of active-site residues\citep{Simonson2002,Nielsen2003,Benkovic2003}. However, empirical force fields are unable to describe the full electronic reorganization in enzyme active sites arising from bond making/breaking and charge transfer along short hydrogen bonds. In addition, most force fields do not include parametrization for transition metals or non-standard ligands. For these systems, single-point electronic structure calculations and geometry optimizations (i.e. quenching to the local 0 K structure) have been highly useful in determining bonding topologies, identifying transition states and intermediates, as well as predicting reaction kinetics \citep{Gao2002,Benkovic2003,Siegbahn2006}. However, one must go beyond these snapshot-based methods to determine the full electron density redistribution as the enzyme's structure fluctuates, which can lead to bond cleavage and proton movement.

\textit{Ab initio} molecular dynamics (AIMD) simulations evolve the nuclei using forces generated from the instantaneous electronic structure obtained at each time step, which allows coupling between nuclear motion and electronic rearrangement. These simulations treat the electrons quantum mechanically and the nuclei classically. However, if light atoms are present, nuclear quantum effects (NQEs), such as tunneling and zero point energy (ZPE), can also play an important role. For example, kinetic isotope effects of over two orders of magnitude on the enzymes' catalytic rates have been experimentally observed upon substituting hydrogen (H) by deuterium (D) \citep{Sutcliffe2002,Klinman2013} and have been attributed to tunneling. Also, the ZPE in a typical oxygen--hydrogen (O--H) bond is $\sim$5 kcal/mol. As a result, including the ZPE can significantly alter the structure and dynamics of systems containing short hydrogen bonds, which are commonly observed in biomolecules where the protein fold can position groups much closer than typically seen in solution \citep{Cleland1998,Mildvan2002}. To treat NQEs one can perform \textit{ab initio} path integral molecular dynamics (AI-PIMD) simulations, which exactly include the effect of NQEs on static equilibrium properties of a given electronic surface \citep{bern-thir86arpc,marx-parr95jcp,morr-car08prl}. The imaginary time path integral formalism on which PIMD simulations are founded also forms the basis for the approximate centroid molecular dynamics (CMD) \citep{cao-voth94jcp,Jang1999} and ring polymer molecular dynamics (RPMD) \citep{crai-mano04jcp,Habershon2013} approaches to obtain quantum dynamics.

Performing AIMD simulations for enzyme active sites has traditionally been a formidable computational task because electronic structure calculations must be performed at each time step. Due to the need to perform many thousands of electronic structure calculations for each picosecond evolved, these simulations are typically performed using density functional theory (DFT) to generate the wave function, although cheaper semiempirical methods such as density functional tight binding have also been employed when longer time scales are required, albeit at a loss in accuracy \citep{Riccardi_2006,Gaus2014}. However, recent theoretical and algorithmic advances and novel computer processing architectures have greatly accelerated these simulations and allowed for trajectories of hundreds of picoseconds or even nanosecond time scales for systems on the order of 60--300 atoms. In particular, electronic structure calculations have been accelerated by the development of linear scaling \citep{linear_scaling_1999, linear_scaling_2012,siesta,conquest,onetep} DFT methods and the advent of codes that use the massively parallel stream processing capabilities of graphical processing units (GPUs) \citep{Ufimtsev_2008, Ufimtsev_2009a}. The former methods take advantage of the locality of the density matrix as well as the fact that both the Coulomb/Hartree and DFT exchange-correlation energies are functions of the local spin density, which allows for a linear growth in computational cost when the system size increases. An example of the latter, which will be the major focus of this article, is the TeraChem electronic structure program, which uses GPUs to accelerate the computation of the electronic wave function and has demonstrated speedups of over 100 fold compared to CPU-based codes \citep{Ufimtsev_2008, Ufimtsev_2009a, Ufimtsev_2009b, Isborn_2011}. This speedup has enabled \textit{ab initio} energy calculations on both the ground and excitepd states of systems containing thousands of atoms, including polypeptides and proteins \citep{Ufimtsev_2008, Ufimtsev_2009a, Ufimtsev_2009b, Isborn_2011, Kulik_2012}.

AI-PIMD simulations have typically required about 2 orders of magnitude more computational cost than the corresponding AIMD simulations. A large amount of this additional cost arises from the need to make many replicas of the system, each of which requires a separate electronic structure calculation. Hence a standard PIMD implementation requires 30--50 electronic structure calculations to evolve a single time step for a typical hydrogen containing system at room temperature \citep{Wallqvist1985,bern-thir86arpc,mark-mano08jcp}. In addition, the PIMD Hamiltonian contains high-frequency motions that limit the time step that can be employed and do not efficiently sample the full phase space if used directly. However, recent developments have significantly alleviated these issues, reducing the computational overhead for including NQEs in simulations \citep{mark-mano08jcp,mark-mano08cpl,ceri+10jcp,ceri-mano12prl} and making it more efficient to extract isotope effects on free energy changes from these simulations \citep{Vanicek2007,ceri-mark13jcp,Marsalek2014}.

Here, we briefly review some considerations when performing electronic structure calculations of large biological systems with a particular emphasis on the GPU-accelerated TeraChem code. We then outline how these electronic structure developments have recently been combined with the latest PIMD algorithms to allow AI-PIMD simulations, which include nuclear and electronic quantum effects, to be performed for enzyme active sites. 

\section{Electronic quantum effects in biological systems}

Electronic structure codes that are written to take advantage of GPUs, such as TeraChem, have made it possible to perform DFT single-point energy calculations, geometry optimizations, and AIMD simulations on quantum mechanical (QM) regions of many hundreds to thousands of atoms. The ability to now perform such large calculations have provided a number of insights into biological processes. For example, large-scale geometry optimizations showed that changes in protein structure upon mutation are directly correlated with enzymatic methyl transfer efficiency \citep{Zhang_2015}. In addition, AIMD simulations have been used to examine charge transfer and polarization in the BPTI protein \citep{Ufimtsev_2011}, discover new pathways for glycine synthesis from primitive compounds \citep{Wang_2014}, and determine amorphous indium phosphide nanostructures \citep{Zhao_2015}.

However, calculations of large systems have also highlighted deficiencies in DFT. In particular, semilocal DFT methods lead to large size-dependent errors and nonlocal exact exchange is necessary to fix these errors. As nonlocal exact exchange is extremely difficult to include in linear-scaling DFT methods, parallel stream processing computer hardware such as GPUs and software advances like those implemented in TeraChem are essential for large-scale exact exchange DFT calculations. While TeraChem is not linear scaling with the number of basis functions, $N$, over a large range of practical system sizes it achieves scaling of around $N^{1.5}$, which is considerably better than the formal DFT scaling of $N^3$ \citep{Luehr2016}. In this section we discuss the choice of basis set, DFT exchange correlation functional, and QM region in the hybrid quantum mechanics/molecular mechanics (QM/MM) calculations for biological systems. 

\subsection{Basis sets}
The basis set describes the electronic wave functions and must balance the accuracy of a given property with computational cost in the quantum chemistry calculation. For example, calculating gas-phase bond dissociation energies with a high level of theory to sub-kcal/mol convergence may require approaching the complete basis set limit \citep{Ochterski_1996, Henry_2002, Haworth_2002}. However, for very large QM calculations, using such a large basis set is often not necessary, and including diffuse functions may lead to difficulties converging the wave function. In many condensed phase applications one is concerned with properties that are less sensitive to basis set effects, e.g. transfer of a proton between similar chemical groups. For such geometry based properties, smaller (e.g. double zeta) basis sets can often be sufficient to obtain the desired accuracy \citep{Wang2004}. Figure \ref{fig:conv_basis} shows the potential energy profile for moving a proton along the active-site hydrogen bond network in an enzyme KSI$^{D40N}$. Except for 6-31G, all the other basis sets produce potential energy curves in quantitative agreement with that predicted using the largest basis set aug-cc-pVDZ.

\begin{figure}[H]
\begin{center}
\includegraphics[width=0.6\textwidth]{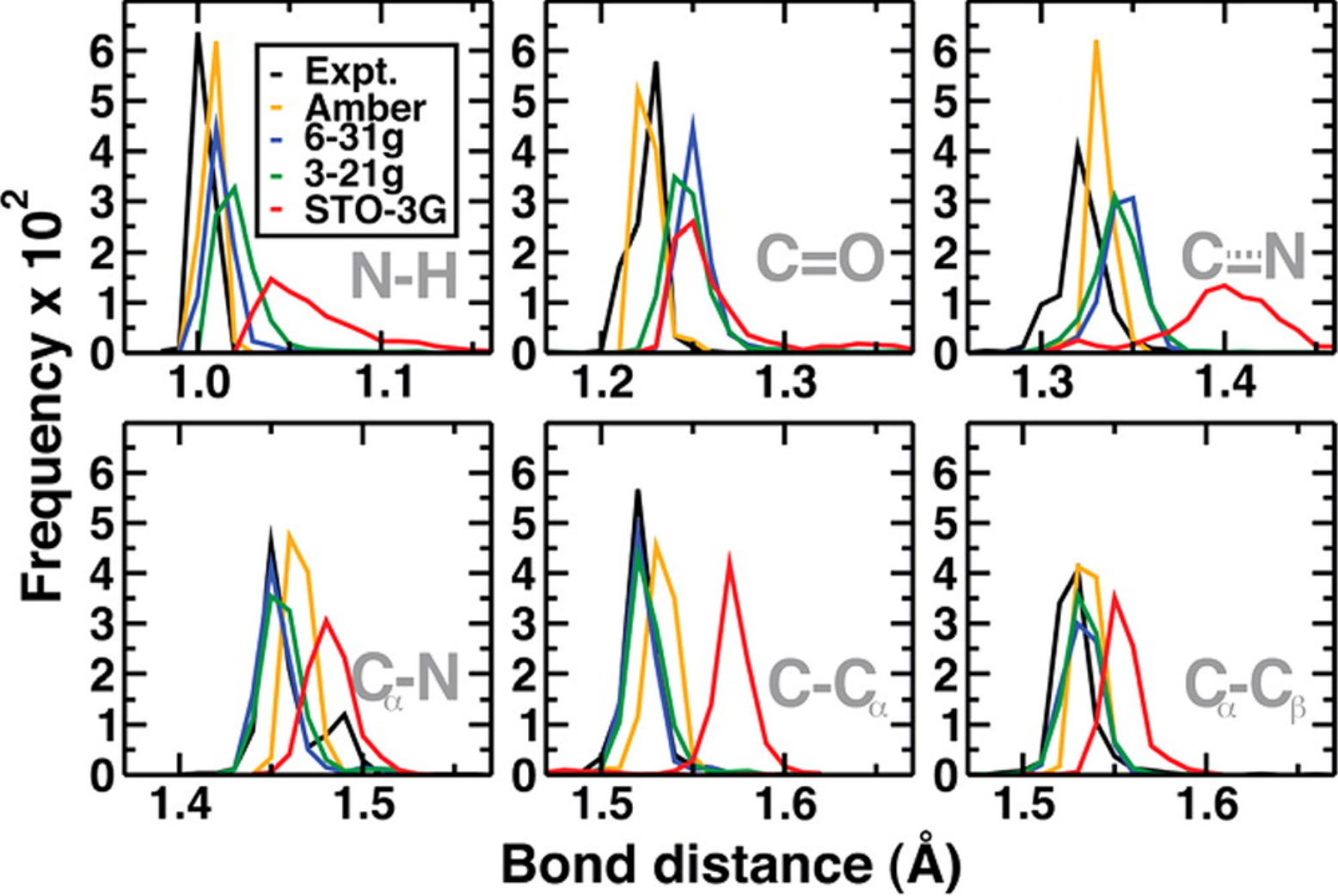} 
\end{center}
\caption{Distributions of bond length for experimental structures (black) compared to those obtained from structural optimization with the classical AMBER force field (orange), $\omega$PBEh/STO-3G (red), $\omega$PBEh/3-21G (green), and $\omega$PBEh/6-31G (blue) for a collection of 58 proteins. (Reproduced from reference \citep{Kulik_2012} with permission from American Chemical Society.) } \label{f:basis} 
\end{figure}

Figure~\ref{f:basis} shows results from Kulik et al., who performed DFT structural optimizations of 58 proteins using TeraChem and showed that small basis sets such as MINI and STO-3G, while computationally efficient, lead to significant errors in predicted properties such as bond lengths and structural clashes as compared to the experimental crystal structures \citep{Kulik_2012}. This problem can be alleviated by using a double-zeta basis set such as 3-21G, although this basis set is still relatively small and thus is likely subject to basis set superposition errors and may be too inflexible for modeling more subtle changes in polarization. For an accurate description of the electronic wave function and hydrogen bonding, the basis set should include polarization functions on all atoms. For systems with electrons far from the nucleus such as anions or excitepd states, diffuse functions should also be included.  

\subsection{Density functionals and the role of exact exchange}

While DFT is formally exact in principle, owing to the Hohenberg-Kohn and Kohn-Sham theorems \citep{Hohenberg_Kohn, Kohn_Sham_1965}, the functional form of the exchange-correlation energy is unknown and therefore it is necessary to make approximations for practical calculations \citep{Perdew_2009, Burke_2012}. DFT methods in practice often approximate the electron exchange and correlation interactions by considering only semi-local properties of the electron density. For example, the generalized gradient approximation (GGA) functionals use the local spin density and its gradient at a given point in space \citep{GGA_1986}, while meta-GGA functionals further include the kinetic energy of the density \citep{TPSS_2003}. These semilocal functionals are often quite successful in regions of slowly varying electron density and can also benefit from cancellation of errors in the exchange and correlation functionals. 

Dispersion is important for accurately modeling thermodynamic and substrate binding properties of enzymes. However, as dispersion is a long-range dynamical correlation effect, it is not accounted for in semilocal density functionals \citep{Johnson_2009, Grimme_2011}. A common technique to overcome this deficiency is to add an explicit, empirically parametrized attractive term to the DFT energy to represent interactions between atomic pairs, in which case `-D' is usually appended to the name of the functional \citep{Wu_2001,Elstner_2001,Grimme_2006,Corminboeuf_2008}. These DFT-D approaches require little additional computational cost and often work quite well. Nevertheless, the accuracy of the DFT-D predictions depend on the size of the basis sets and are limited by the atom types included in the parametrization. An alternative, non-empirical approach is to compute the dispersion interaction between atoms based on the exchange-hole dipole model (XDM) \citep{Becke_2005,Johnson_2006,Becke_2007}. This model also adds negligible computational cost to a DFT calculation, but its implementation in geometry optimizations and dynamical simulations is currently not practical because forces obtained from the XDM method have not been widely available.

The approximate exchange term in semilocal DFT methods produces errors in both the electron density and energy, which leads to the derivative discontinuity \citep{perdew1982,Tozer_2003,Kronik-2014,Kronik-2014-b}, the many-electron self-interaction error \citep{Weitao-1998,Weitao-2006} and the delocalization error \citep{Weitao-2008,Weitao-2010,Weitao-2012,kim2013}.
These semilocal density functionals underestimate the energy gap between the highest-occupied and lowest-unoccupied molecular orbitals (HOMO and LUMO) \citep{Weitao-2008, Weitao-2008-c, Weitao-2008-d, Baer-2012}, which leads to severe problems as the size of the QM region increases. We have shown that for molecules in aqueous solution, semilocal DFT methods predict that the band gap approaches zero and eventually the ground state self-consistent field (SCF) calculation can no longer converge if more and more solvent molecules are included in the QM region \citep{Isborn_2013}. Similar difficulties were found by Kulik et al. when performing semilocal DFT calculations on entire proteins \citep{Kulik_2012}. Additionally,  we have recently shown that approximate semilocal exchange functionals generate a size-dependent error in the ionization potential, with the size of the error increasing as the size of the system grows \citep{Whittleton_2015, Sosa_Vazquez_2015}.

Many of these inaccuracies can be improved by incorporating nonlocal exact exchange into the density functional via the generalized Kohn-Sham scheme \citep{Seidl_1996}. Exact exchange is often included in approximate semilocal exchange methods through either global hybrid or range-separated hybrid (RSH) techniques. For global hybrids, such as the B3LYP functional \citep{Becke1993}, a fixed fraction of exact exchange (typically 20--50$\%$) is included for all interelectronic distances. RSH methods vary the amount of exact exchange based on the interelectronic distance $r_{12}\equiv  | \mathbf{r}_{12}| = | \mathbf{r}_{1} - \mathbf{r}_{2} | $, with the error function usually used to make the variation smooth for splitting the short-and long-range parts of the Coulomb operator
\begin{equation}
\frac{1}{r_{12}} = \frac{1-erf({\omega r_{12}})}{r_{12}} + \frac{erf({\omega r_{12}})}{r_{12}}.
\end{equation}
For a long-range corrected (LC) hybrid, full exact exchange is used at long range (the $erf({\omega r_{12}})$ term), while semilocal DFT exchange dominates at short-range (the $1-erf({\omega r_{12}})$ term) to provide good balance with semilocal DFT correlation.  The ratio of local DFT exchange to nonlocal exact exchange is determined by a range separation parameter, $\omega$, given in atomic units of inverse bohrs, $a_0^{-1}$ \citep{gill1996, iikura2001,Hirao-2004,Handy-2004,Savin-2004,Baer-2005,Head-Gordon-2008}. 
The default value of $\omega$ is often between 0.2 and 0.5 $a_0^{-1}$, with smaller values leading to local DFT exchange having a longer range. A system-dependent `optimal' $\omega$ can also be determined by tuning $\omega$ to enforce Koopmans' theorem \citep{Baer-2009-c, Baer-2010, Baer-2010-review, Kronik_2012, Autschbach-2014_ACR}.

\begin{figure} [H]
\begin{center}
\includegraphics[width=0.8\textwidth]{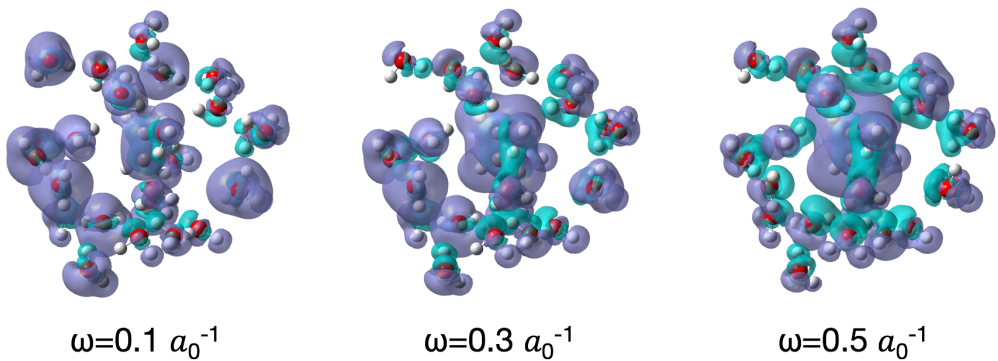} 
\end{center} 
\caption{The DFT error in size-dependent ionization potential is shown by analyzing the neutral and cation SCF density differences for ethene surrounded by 25 water molecules for the LC-BLYP functional with $\omega$=0.1, 0.3, and 0.5 $a_0^{-1}$. The purple isosurface shows where electron density has been removed, and the blue isosurface shows where electron density has been gained. The correct polarization response of the solvent is seen for $\omega$=0.5  $a_0^{-1}$, where the water molecules polarize their electron density towards the cationic solute. Smaller $\omega$ values show the solvent being incorrectly partially ionized due to the DFT delocalization error present when using approximate exchange based on the local spin density.} \label{f:water_omega}
\end{figure}

Long-range corrected hybrid functionals have the correct asymptotic limit for the exchange interaction, which often yields improved ionization potentials, band gaps, and excitation energies compared to experiment \citep{Baer-2011, Baer-2011-c,Baer-2012, Kronik_2012,Baer-2012-b, Autschbach-2014_ACR}, as well as improved convergence of the SCF ground state for systems containing large QM regions \citep{Kulik_2012, Isborn_2013}. However, although all long-range corrected hybrid functionals have the correct asymptotic behavior of the exchange interaction, care must be taken to ensure that the range-separation parameter $\omega$ is large enough to fix the deficiencies of using approximate semilocal exchange at short range. We have recently shown that the choice of $\omega$ is key in determining the correct physical polarization in response to electron ionization. 
For example, for the ionization of a solvated ethene molecule, if $\omega$ is too small (below $\sim$0.4 $a_0^{-1}$), the surrounding water molecules will unphysically contribute some of their own electron density to the ionization process, while for a larger $\omega$ they will correctly polarize in response to the electron being removed from the solute \citep{Sosa_Vazquez_2015}. This can be seen in the density differences for the neutral and cation systems shown in Figure~\ref{f:water_omega}.  Thus the correct choice of functional, and particularly the treatment of exchange in the functional, is crucial for correctly describing processes that are key in enzymatic reactions such as polarization and charge-transfer. Hence, in addition to including dispersion through methods mentioned above, we recommend a long-range corrected hybrid functional such as LC-BLYP, LC-$\omega$PBE, or $\omega$B97, with an $\omega$ value in the range $\omega$=0.4-0.6 $a_0^{-1}$. 

\subsection{QM regions in QM/MM calculations}
When simulating large biomolecules, it is desirable to accurately describe the region of interest, such as an enzyme's active site, using a high-level QM method with a large basis set. Such quantum mechanical treatments are only feasible for a fraction of atoms in biomolecules due to the high computational cost of the QM approach. However, unlike an isolated system in vacuum, the electrostatic environment in the condensed phase strongly polarizes the QM system and therefore it is crucial to include the environmental effects for a correct description of the energy, geometry, and electron density distribution in the QM region.
The hybrid QM/MM approach takes advantage of both the accuracy of the QM method and the speed of the MM method, allowing the MM charges to polarize the QM wave function \citep{Warshel_1976, Friesner_2005}. The coupling between the QM and MM regions has been extensively investigated, see Refs. \citep{Lennartz_2002, Altun_2006} for examples; and for recent reviews of the QM/MM method see Refs. \citep{Riccardi_2006, Lin2006, Hans_Martin_2009, Ranaghan_2010}

To combine large-scale GPU-accelerated quantum chemistry calculations with MM force fields, we have created an extensive TeraChem interface for QM/MM calculations \citep{Isborn_2012}.
In this approach, TeraChem receives the point charges of the MM atoms and performs a QM calculation with the electrostatic embedding method to account for the polarization of the QM region due to the electrostatic MM environment. TeraChem then outputs the QM forces on each atom for the propagation of the QM/MM system. Communication between the QM and MM regions is either through files or the message passing interface (MPI). Although the TeraChem QM/MM interface was originally implemented with the AMBER molecular dynamics program \citep{amber_12}, it is general enough to be used with any program that can accept quantum mechanical forces for molecular dynamics simulations. Indeed, the AI-PIMD discussed in the following section were performed using the TeraChem MPI interface to connect to an in-house PIMD code to perform the dynamics.

\begin{figure} [H]
\begin{center}
\includegraphics[width=0.6\textwidth]{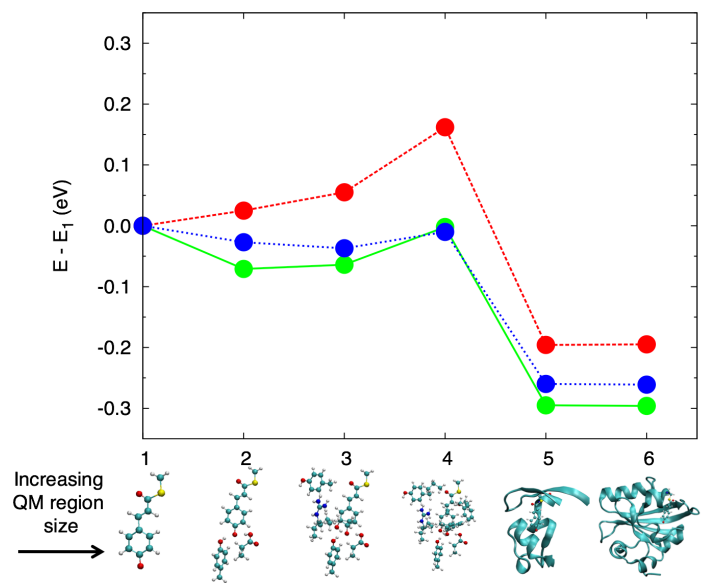} 
\end{center} 
\caption{Dependence of the PYP excitepd state energy on the size of the QM region, as obtained from QM/MM calculations for three different PYP configurations. Very large QM regions are required to generate the converged excitation energies. The y-axis is the difference in the computed excitepd state energy compared to that for QM region 1.} \label{f:pyp} 
\end{figure}

Ideally, the QM region in the QM/MM calculation will be large enough that the results are independent of QM region size. To test how large the QM region should be to compute converged properties, we performed excitepd state QM/MM calculations on photoactive yellow protein (PYP) and computed its absorption spectrum using six different QM regions of increasing size, which range from the photoactive chromophore alone (22 QM atoms, 217 basis functions) to including the entire protein and counter ions (1935 QM atoms, 16827 basis functions) \citep{Isborn_2012}. For all calculations the total system was the same, which included the entire protein, counter ions, and a 32 \AA  ~solvation sphere, and any atoms not included in the QM region were modeled with MM point charges. We found that although the electronic excitation was for the most part localized on the chromophore in all QM/MM calculations, a large QM region (723 QM atoms) was required to reproduce the energy of the largest QM system (the entire protein included in the calculation). Furthermore, the predicted excitation energy did not have a monotonic trend as the size of the QM region increased, but instead jumped around substantially (up by 0.4 eV when going from the 4$^{th}$ to 5$^{th}$ largest QM region) as additional protein residues were included in the QM region, see Figure~\ref{f:pyp}. Overall, when sampling over many snapshots, this dependence of excitation energy on the size of the QM region led to significant differences in the final computed absorption spectrum \citep{Isborn_2012}. 

The slow convergence of a local property, such as an excitation energy, with the size of the QM region suggests that the electrostatic interactions at the QM/MM boundary can cause large changes in the QM electron density. Therefore the QM region should be as large as possible to push this boundary away from the active site to obtain an accurate QM density. Others have arrived at similar conclusions about the large size required for the QM region, citing problems with charges at the QM/MM junction\citep{Sumowski_2009,Lopez-Canut_2009,Hu_2011,Fox_2011,Flaig_2012}.

\section{Incorporating nuclear quantum effects in AIMD simulations}

The interplay between nuclear and electronic quantum effects can dramatically alter the structure and dynamics of hydrogen bonded systems. It has been shown that the distance between the donor and acceptor heavy atoms, R, plays a crucial role in determining the proton behavior in a hydrogen bond \citep{Perrin1997,McKenzie2012}. In particular, when R is around or below 2.7 \AA, the influence of NQEs becomes significant. For example, in liquid water, which has an average O--O distance of around 2.8 \AA, the quantum nature of the hydrogen bonds leads to transient proton excursion events in which a proton is closer to the hydrogen bond acceptor than donor oxygen \citep{ceri+13pnas,Wang2014}.

Despite the importance of NQEs, most molecular simulations are performed treating the nuclei as classical particles. The path integral formalism of quantum mechanics provides an efficient way to incorporate nuclear quantum effects in molecular simulations \citep{feyn-hibb65book}. By combining the path integral methods and on-the-fly electronic structure calculations, one can perform AI-PIMD simulations to examine the nuclear and electronic quantum fluctuations in chemical and biological systems. Below we provide the basic concepts, implementation, and examples for AI-PIMD simulations.

\subsection{Path integral molecular dynamics}
PIMD simulations allow the exact inclusion of NQEs on static equilibrium properties for a given electronic surface by exploiting the isomorphism between the quantum partition function of a quantum mechanical system and the classical partition function of a ``ring polymer", which is a cyclic structure containing multiple copies of the classical system with adjacent copies (beads) connected by harmonic springs \citep{bern-thir86arpc}.

For a system of $N$ classical particles of masses $m_i$ the Hamiltonian is
\begin{equation}
H = \sum_{i=1}^N \frac{{\bf p}_i^2}{2m_i} + V({\bf r}_1,\ldots,{\bf r}_N),
\end{equation}
where in AIMD the potential energy $V({\bf r}_1,\ldots,{\bf r}_N)$ is obtained from an electronic structure calculation. The path integral expression for the partition function for this Hamiltonian is
\begin{equation}
\label{eq:PIMD}
Q_P = \frac{1}{(2\pi\hbar)^{f}} \int d^f{\bf p} \int d^f {\bf r} \
e^{-\beta_{P} H_P({\bf p}, {\bf r})},
\label{eq:Z_P}
\end{equation}
where $f=3NP$, $P$ is the number of beads in the ring polymer, ${\bf p} \equiv \{{\bf p}_i^{(j)}\}_{i=1 \ldots N}^{j=1 \ldots P}$,  ${\bf r} \equiv \{{\bf r}_i^{(j)}\}_{i=1 \ldots N}^{j=1 \ldots P}$ and $\beta_P=\frac{\beta}{P}=\frac{1}{k_B T_P}$. The effective temperature, $T_P=PT$, under which each bead evolves is thus $P$ times higher than the physical temperature, allowing the copies of the system to sample areas of phase space that are not accessible to a classical particle. The PIMD Hamiltonian $H_P({\bf p},{\bf r})$ is,
\begin{equation}
\label{eq:HP}
H_P({\bf p},{\bf r}) = \sum_{j=1}^P \left( \sum_{i=1}^N {\frac{|{\bf p}_{i}^{(j)}|^2}{2 m_i}} +                                \frac{1}{2}m_i\omega_P^2 |{\bf r}_{i}^{(j)}-{\bf r}_{i}^{(j-1)}|^2 \right) +  V({\bf r}_1^{(j)},\ldots,{\bf r}_N^{(j)}),
\end{equation}
where $\omega_P=1/{\beta_P}\hbar$ and cyclic boundary conditions, $ j + P \equiv j $ are implied. $H_P$ corresponds to the Hamiltonian of a classical ring polymer, in which adjacent beads are linked by harmonic springs with a force constant of $m_{i} \omega_P^2$. In PIMD simulations the system is evolved using Hamilton's equations of motion from this Hamiltonian \citep{bern-thir86arpc}. Note that PIMD simulations utilize the momentum term in Eq. \ref{eq:HP} to sample phase space and hence the dynamics generated from this Hamiltonian are not exact quantum dynamics. 

Static equilibrium properties of the quantum mechanical system can be exactly calculated as 
\begin{equation}
    <A>=\frac{ \int d^f{\bf p} \int d^f{\bf r} ~Ae^{-\beta_P H_P} }{\int d^f{\bf p} \int d^f{\bf r} ~e^{-\beta_P H_P} }.
\end{equation}

\subsection{Accelerating PIMD convergence using generalized Langevin equations}
Eq. \ref{eq:PIMD} is exact when $P\rightarrow\infty$. In practice, a finite number of replicas $P$ is used to converge the properties to the desired accuracy. A useful indication as to the number of replicas is $P > \beta\hbar\omega_{max}$ where $\omega_{max}$ is the highest frequency present in the system. This criteria can be understood as stating that the role of $P$ is to increase the effective thermal energy of each replica ($Pk_BT$) to be larger than the energy level spacings in the system ($\hbar\omega_{max}$ for a harmonic oscillator), i.e. allowing each replica to approach the classical limit \citep{Wallqvist1985,mark-mano08jcp}. For example, the number of replicas used for convergence is typically $P \ge 32$ for a hydrogen-containing system at 300 K. Since each replica of the system requires an {\it ab initio} calculation to evaluate its energy and forces, the computational cost of PIMD simulations is therefore at least $P$ times higher than the corresponding classical simulation.

Computational overhead due to the convergence requirement can be significantly reduced using a generalized Langevin equation (GLE) thermostat \citep{ceri+10jctc,ceri+11jcp,ceri-mano12prl}. 
By coupling the path integral system to a GLE thermostat (the PI+GLE approach), one can tune the correlated noise such that convergence of the potential energy, $\langle V \rangle$, to its quantum mechanical expectation value can be achieved with considerably fewer path integral beads \citep{ceri+10jctc,ceri+11jcp}. However, the PI+GLE approach does not guarantee fast convergence of the quantum kinetic energy, $\langle T \rangle$, as the centroid virial estimator for the kinetic energy (Eq. \ref{eq:QKE}) \citep{herm-bern82jcp,cao-bern89jcp}  involves correlations between the beads and the ring-polymer centroid. To allow rapid convergence of the kinetic energy, Ceriotti and coworkers have recently introduced the PIGLET method, which enforces the necessary correlations to accelerate the convergence of the centroid virial kinetic energy estimator \citep{ceri-mano12prl}. The PIGLET approach allows the convergence of thermodynamic properties and isotope effects that depend on the kinetic energy (see discussions in the next section) using $P=6$ for hydrogen-containing systems at 300 K. This reduces the computational cost by over 5 fold compared to a typical PIMD simulation  \citep{ceri-mano12prl,ceri+13pnas}. Parameters for the GLE thermostat can be conveniently generated using the GLE4MD input generator \citep{gle4md}.  

\subsection{Extracting thermodynamic isotope effects from PIMD simulations} 
Isotope substitution techniques have been widely used in fields ranging from molecular biology to atmospheric chemistry \citep{Kohen2005}. Computer simulations that accurately and efficiently predict isotope effects form an important complement to experiments and have enabled detailed analysis of the reaction mechanism and thermodynamic equilibrium in a wide variety of chemical and biological processes \citep{Kohen2005}.

Since isotopes of the same element have almost identical electronic structures, the change in structural and thermodynamic properties due to isotope substitution arise entirely from the quantum mechanical nature of the nuclei. The central quantity for calculating these thermodynamic isotope effects is the free energy change upon isotope substitution, which can be computed exactly using PIMD simulations for a given electronic surface. For example, the free energy change upon exchanging D for H in a given system $k$, $\Delta A_k$, is directly related to the quantum kinetic energies of the isotopes by the thermodynamic integration \citep{ceri-mark13jcp}
\begin{equation}
\label{eq:freeenergy}
    \Delta A_k=\int_{m_H}^{m_D}{d\mu\frac{\langle T_k(\mu)\rangle}{\mu}}.
\end{equation}
Here $m_H$ and $m_D$ are the masses of H and D, respectively. $T_k(\mu)$ is the quantum kinetic energy of a hydrogen isotope of mass $\mu$ and can be determined from AI-PIMD simulations using the centroid virial estimator \citep{herm-bern82jcp,cao-bern89jcp},
\begin{equation}
\label{eq:QKE}
    \langle T_k(\mu)\rangle=\frac{3}{2\beta}+\frac{1}{2P}\sum_{j=1}^{P}\langle (\mathbf{r}^{(j)}-\mathbf{\bar{r}})\cdot\frac{\partial V}{\partial\mathbf{r}^{(j)}} \rangle,
\end{equation}
where V is the potential energy of the system and $\mathbf{\bar{r}}=\sum_{j=1}^{P}{\mathbf{r}^{(j)}}/P$ is the centroid position.

From Eq. \ref{eq:freeenergy}, one can in principle perform multiple PIMD simulations with varying $\mu$ and obtain $\Delta A_k$ by carrying out the integration. This process can be greatly accelerated by using the free energy perturbation (FEP) method \citep{ceri-mark13jcp}, which allows one to extract $\langle T_k(\mu) \rangle$ from a single simulation of the most abundant isotope of hydrogen. From the FEP approach \citep{ceri-mark13jcp}, 
\begin{equation}
    \langle T_k(\mu)\rangle=\frac{\langle T_k(m_H)e^{-h(\mu/m_H;\mathbf{r})}\rangle_{m_H}}{\langle e^{-h(\mu/m_H;\mathbf{r})}\rangle_{m_H}}
\label{eq:FEP}
\end{equation}
with
\begin{equation}
    h(\alpha;\mathbf{r})=\frac{(\alpha-1)\beta m_H\omega_P^2}{2P}\sum_{j=1}^P(\mathbf{r}^{(j)}-\mathbf{r}^{(j+1)})^2.
\end{equation}

In the following, we use $pK_a$ as an example and demonstrate how one can efficiently extract thermodynamic isotope effects from PIMD simulations. Let us consider the side-chain $pK_a$ of an amino acid X in an enzyme upon H/D substitution. In H$_2$O, this $pK_a$ corresponds to the chemical equilibrium where the neutral amino acid, $XH_{EnzH}$, dissociates into the anion $X^-_{EnzH}$ and the proton $H^+$ (shown in the scheme below). In D$_2$O, we denote the neutral and ionized species of X as $XD_{EnzD}$ and $X^-_{EnzD}$, respectively. Note that in D$_2$O all labile protons in the enzyme are exchanged to D, and thus $X^-_{EnzH}$ and $X^-_{EnzD}$ are not necessarily identical. These chemical reactions can be connected using the thermodynamic cycle:
\begin{figure}[H]
\begin{center}
    \includegraphics[height=2cm]{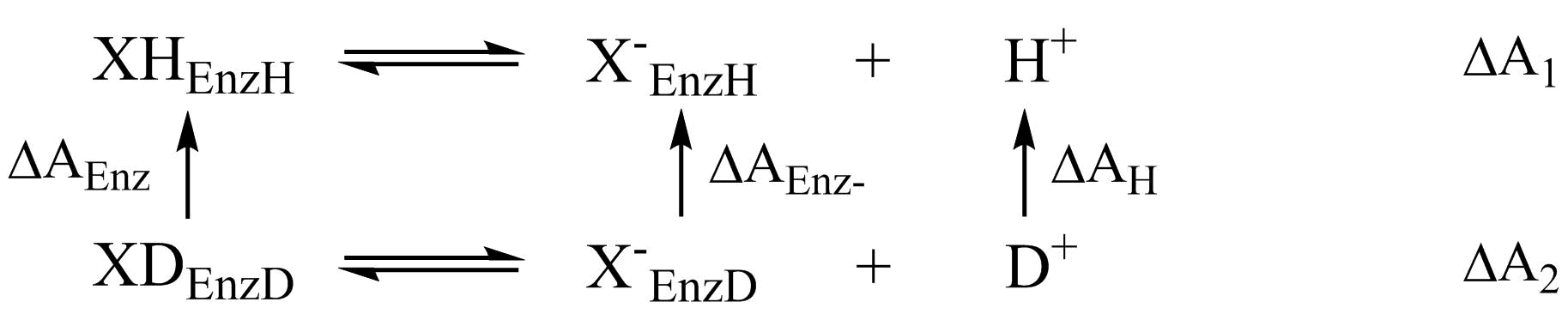}
\end{center}
\end{figure}
Assuming the free energy changes of the aforementioned dissociation processes are $\Delta A_1$ and $\Delta A_2$, then the $pK_a$ isotope effect is
\begin{equation}
    \Delta pK_a^{Enz}\equiv pK_a^{EnzD}-pK_a^{EnzH}=\frac{\Delta A_2-\Delta A_1}{2.303k_BT}=\frac{\Delta A_{Enz}-\Delta A_{Enz-}-\Delta A_H}{2.303k_BT}.
\label{eq:enz}
\end{equation}
$\Delta A_{Enz}$ and $\Delta A_{Enz-}$ are the free energy changes upon converting $XD_{EnzD}$ and $X^-_{EnzD}$ to $XH_{EnzH}$ and $X^-_{EnzH}$, respectively, and can be efficiently calculated from PIMD simulations using Eq. \ref{eq:freeenergy} along with the FEP method (Eq. \ref{eq:FEP}).

To obtain the absolute value of $\Delta pK_a^{Enz}$ one also needs $\Delta A_H$, which is the free energy change upon converting D to H in aqueous solution. It can be computed from PIMD simulations of protons in liquid water using the same level of QM description as that for the enzyme. However, it is often computationally costly to properly sample the solvent configurations around the proton in order to determine an accurate $\Delta A_H$. Alternatively, a useful metric is to compare $\Delta pK_a^{Enz}$ to a reference state such as the $pK_a$ isotope effect of the amino acid X in aqueous solution, $\Delta pK_a^{Sol}$. This results in an excess isotope effect, $\Delta\Delta pK_a\equiv\Delta pK_a^{Enz}-\Delta pK_a^{Sol}$, which represent the extra amount of NQEs arising from the unique enzyme environment as compared to aqueous solution. 

Similar to the enzyme case, $\Delta pK_a^{Sol}$ can be computed from the thermodynamic cycle
\begin{figure}[H]
\begin{center}
    \includegraphics[height=2cm]{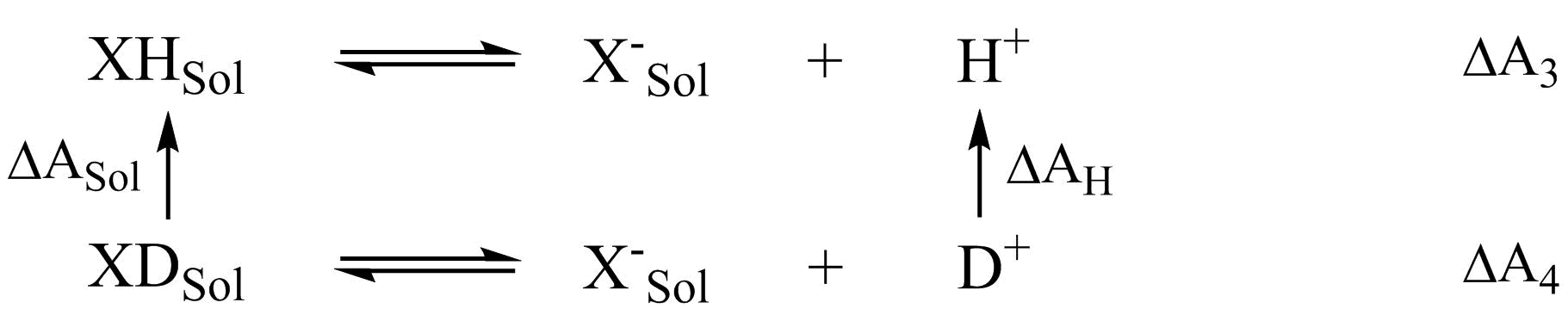}
\end{center}
\end{figure}
$XH_{Sol}$ and $XD_{Sol}$ denote the neutral amino acid X in H$_2$O and D$_2$O, respectively, while $X^-_{Sol}$ represent the ionized species. Therefore,   
\begin{equation}
    \Delta pK_a^{Sol}=pK_a^{SolD}-pK_a^{SolH}=\frac{\Delta A_4-\Delta A_3}{2.303k_BT}=\frac{\Delta A_{Sol}-\Delta A_H}{2.303k_BT}.
\end{equation}
Comparing $\Delta pK_a^{Enz}$ and $\Delta pK_a^{Sol}$ makes the $\Delta A_H$ term cancel, and thus the excess isotope effect is
\begin{equation}
\label{eq:Excess}
    \Delta\Delta pK_a=\Delta pK_a^{Enz}-\Delta pK_a^{Sol}=\frac{\Delta A_{Enz}-\Delta A_{Enz-}-\Delta A_{Sol}}{2.303k_BT}.
\end{equation}
In this way, $\Delta\Delta pK_a$ can be extracted from PIMD simulations by computing the free energy changes upon isotope substitution in the enzyme and solution environment.

\subsection{Electronic and nuclear quantum fluctuations in biological hydrogen bond networks} \label{sec:ksi}
Combing the technical advances introduced earlier, we now demonstrate how one can employ AI-PIMD simulations to elucidate the interplay between electronic and nuclear quantum effects in enzymes. We have recently studied the behavior of an active-site hydrogen bond network in an enzyme mutant KSI$^{D40N}$ by interfacing with TeraChem and taking advantage of the PIGLET algorithm \citep{ceri-mano12prl} and the FEP method \citep{ceri-mark13jcp}. As the majority of computational cost comes from \textit{ab initio} force evaluations on the $P$ ring-polymer beads, we have parallelized these calculations  across GPUs with one GPU dedicated for each bead. These methods allow us to achieve a simulation speed that is three orders of magnitude higher than existing AI-PIMD approaches. In addition, in TeraChem the speed of electronic structure evaluations scales almost linearly with the number of GPUs. Therefore, one can further speedup the simulations by assigning one node for each bead and using multiple GPUs on each node to carry out electronic structure calculations.

\begin{figure} [H]
\center
    \includegraphics[height=5cm]{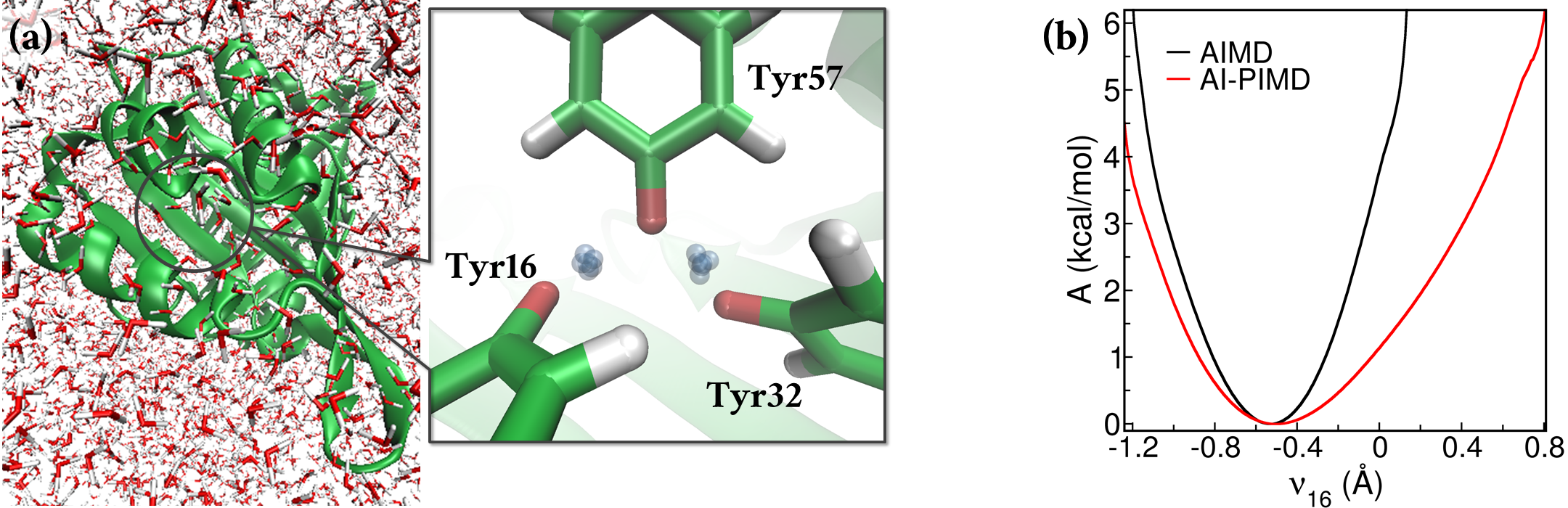}
    \caption{(a) A snapshot of KSI$^{D40N}$ from the AI-PIMD simulation. Its active-site hydrogen bond network enlarged, in which green, red and white represent carbon, oxygen and hydrogen atoms. The blue spheres are the full ring-polymer representation of the protons. For clarity, all the other atoms are shown as their centroids. (b) The free energy surface of the proton movement along $\nu_{16}$, as calculated from AIMD and AI-PIMD simulations.}
    \label{fig:KSI_structure}
\end{figure}

In KSI$^{D40N}$, the side chain groups of the active-site residues form a triad structural motif, in which residue Tyr57 sits at the center and forms two hydrogen bonds with the hydroxyl groups of residues Tyr16 and Tyr32 (Figure \ref{fig:KSI_structure}(a)). In several high-resolution crystal structures \citep{Ha2000,Sigala2013} the hydrogen bonded O57--O16 and O57--O32 distances are found to be around 2.6 \AA, noticeably shorter than those commonly observed in hydrogen bonded systems such as liquid water.

\begin{figure}[H]
\begin{center}
    \includegraphics[height=6cm]{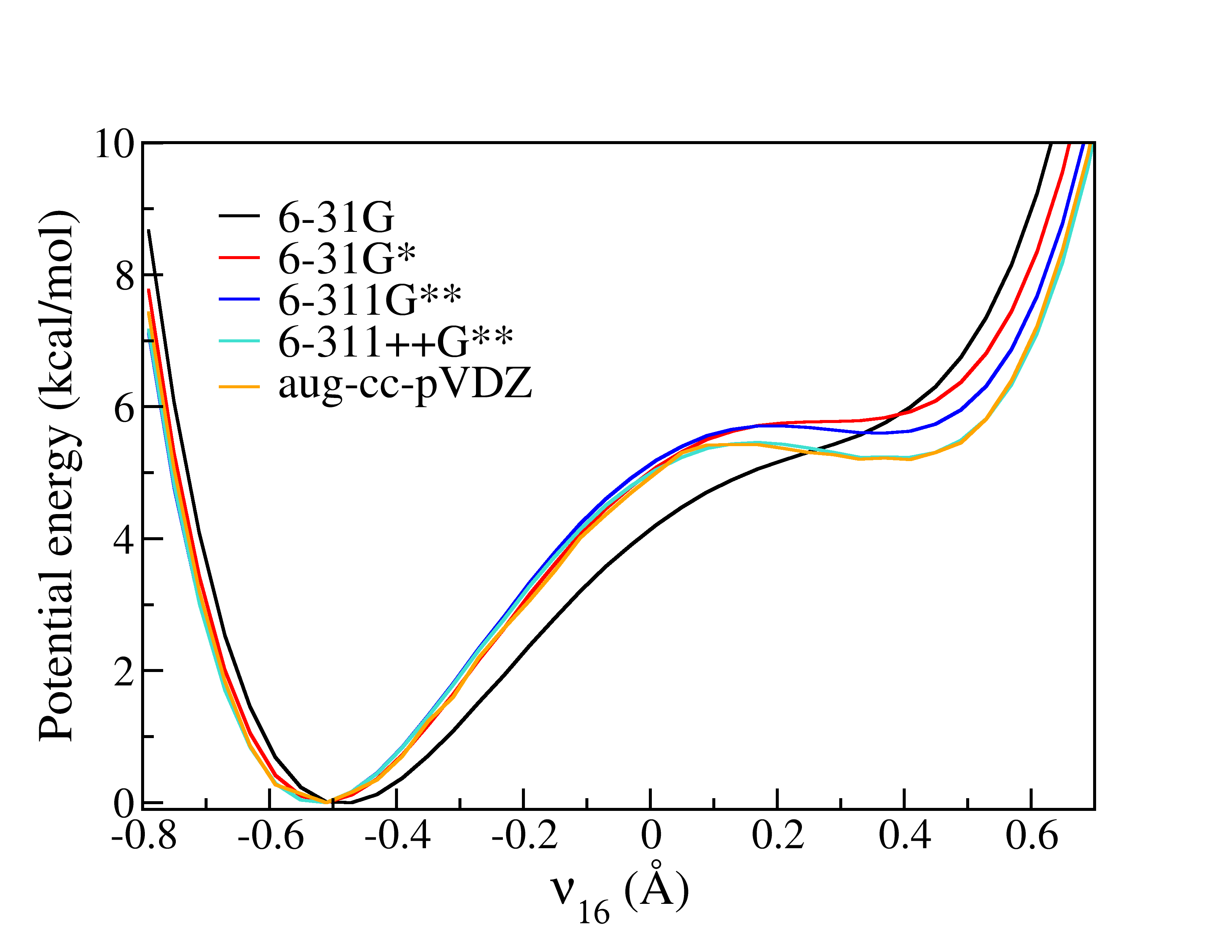}
\end{center}
    \caption{Potential energy profile for proton transfer in the active-site hydrogen bond network of KSI$^{D40N}$, as calculated at the B3LYP-D3 level \citep{Becke1993,grim+10jcp} using five different basis sets. The proton transfer coordinate $\nu_{16}=d_{O16,H16}-d_{O57,H16}$ and $\nu_{16} \ge 0$ represents a proton transfer from residue Tyr16 to Tyr57.}
    \label{fig:conv_basis}
\end{figure}

\begin{figure} [H]
\center
    \includegraphics[height=8cm]{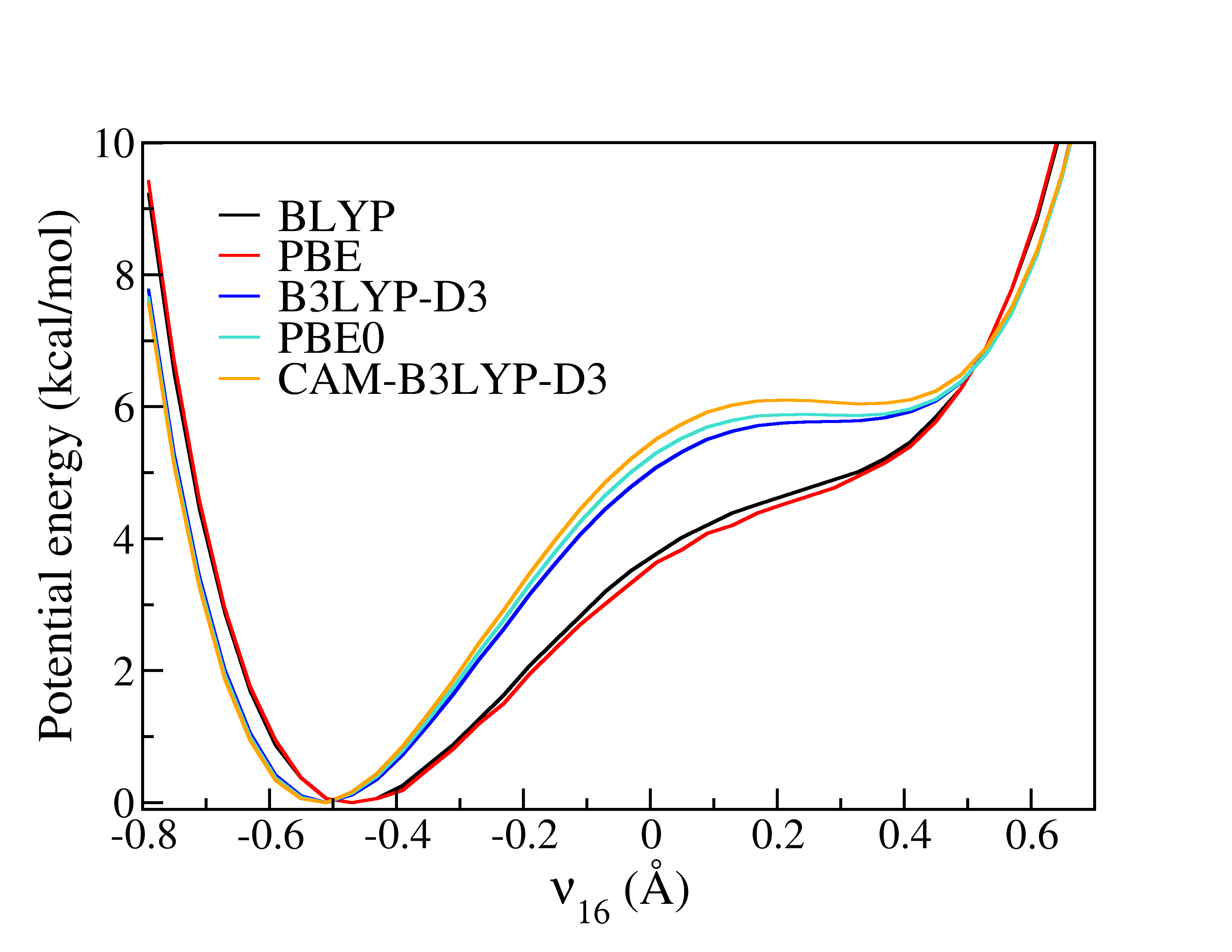}
    \caption{Potential energy profile for proton transfer in KSI$^{D40N}$, as obtained using five density functionals and the 6-31G* basis set.}
    \label{fig:conv_method}
\end{figure}

In order to choose the proper DFT functional and basis set for the QM region, we have calculated the potential energy profiles for proton transfer within the tyrosine triad. 
We define the proton transfer coordinate as $\nu_{16}=d_{O16,H16}-d_{O57,H16}$, where $d_{Oi,Hj}$ is the distance between the oxygen O of residue $i$ and hydrogen H of residue $j$ (Figure \ref{fig:KSI_structure}(a)), and a quantity $\Delta E_{\nu=0}$ which represents the energy required to move the proton from its equilibrium position to the perfectly shared position, $\nu_{16}=0$.
As shown in Figure \ref{fig:conv_method}, all density functionals tested generate results that qualitatively agree with each other, although GGA functionals BLYP and PBE underestimate $\Delta E_{\nu=0}$ by about 1.5 kcal/mol. Hybrid functionals B3LYP and PBE0 produce almost identical potential energy profiles, with the CAM-B3LYP RSH method in excellent agreement with the global hybrids. Dispersion corrections (D3 in this case) lead to very little change. Our basis set studies in Figure \ref{fig:conv_basis} demonstrate that 6-31G* is the smallest basis set that is able to produce the correct energy profile for proton transfer. Based on these tests, we choose to treat the QM region at the B3LYP-D3 level \citep{Becke1993,grim+10jcp} with the 6-31G* basis set.

In our AI-PIMD simulations, the QM region includes the tyrosine triad (47 atoms and 174 electrons) and is treated quantum mechanically in both the electronic and nuclear degrees of freedom. We have shown that increasing the size of the QM region by incorporating nearby residues does not significantly change the potential energy profile for proton transfer \citep{Wang2014a}. As environmental fluctuations are indispensable for a correct description of the QM region, especially the O--O distances, we include the rest of the protein as well as solvent molecules and counter-ions in the MM region (over 52,000 atoms), which is described using the AMBER03 \citep{JCC_2003} force field and the TIP3P water model \citep{jorg+83jcp}.
All bonds across the QM/MM boundary are capped with hydrogen link atoms in the QM region, which are constrained along the bisecting bonds and do not interact with the MM region.
The total interactions in the QM/MM system include 1) forces between QM atoms, 2) forces between MM atoms and 3) electrostatic and van der Waals interactions between QM and MM atoms. Our simulation program obtains forces within the QM region and the electrostatic QM/MM interactions through an interface with TeraChem \citep{Ufimtsev_2009b,Isborn_2012}, while the forces within the MM region and the Lennard-Jones QM/MM interactions are acquired via an MPI interface with the LAMMPS molecular dynamics package \citep{plim95jcp}. A snapshot of the QM/MM system is shown in Figure \ref{fig:KSI_structure}(a).

The impact of NQEs can be understood by comparing AIMD and AI-PIMD simulations of KSI$^{D40N}$. Figure \ref{fig:KSI_structure}(b) shows that NQEs significantly decrease the free energy required to move the proton from Tyr16 to Tyr57 ($\nu_{16} \ge 0$), thus allowing the proton to be quantum mechanically delocalized in the active-site hydrogen bond network. This is manifested by the wide spread of the protons' ring polymer beads, which represents their position uncertainty (Figure \ref{fig:KSI_structure}(a)).
This quantum delocalization leads to a 10,000 fold increase in the acidity dissociation constant of Tyr57 and a $\Delta\Delta pK_a$ of 0.50 compared to tyrosine in solution, in excellent agreement with experiment \citep{Wang2014a}.

Due to the short R in the active-site hydrogen bond network, the ZPE possessed by the O--H bonds and $\Delta E_{\nu=0}$ are both about 5 kcal/mol (Figure \ref{fig:conv_method}), which facilitates proton movements along $\nu_{16}$. Hence the interplay between nuclear and electronic quantum effects qualitatively and quantitatively changes the proton behavior from classical hydrogen bonding to quantum delocalization. 

\section{Outlook}
With the development of new algorithms and their combination with novel computer streaming architectures, path integral methods and AIMD simulations can now be combined to elucidate the movements of electrons and light nuclei in unprecedented detail during biological processes. In addition, recent developments will accelerate methods such as CMD \citep{cao-voth94jcp,Jang1999} and RPMD \citep{crai-mano04jcp,Habershon2013} to obtain approximate quantum dynamics. These methods are not amenable to acceleration using the PIGLET approach. However, very recently it has been shown how the ring-polymer contraction (RPC) approach \citep{mark-mano08jcp,Fanourgakis2009}, which exploits the separation of the forces acting on each copy of the system (Eq. \ref{eq:HP}) into rapidly and slowly varying parts, can be extended to AI-PIMD and RPMD simulations \citep{Marsalek2016,Kapil2016}. 

As was illustrated in Sec.~\ref{sec:ksi}, AI-PIMD simulations can already provide novel insights into the structure and functional roles of short hydrogen bonds (R $\le$ 2.6 \AA) in biological systems. Another area where AI-PIMD simulations may be able to offer important insights is the study of proton coupled electron transfer (PCET) reactions, which play a fundamental role in photosynthesis, respiration, and enzymatic catalysis  \citep{Chang2004,Cukier2004,Hammes-Schiffer2010}. A comprehensive understanding of these biological processes requires simulation techniques that properly describe the quantum mechanical nature of the electrons and the light nuclei. With the recent advent of  efficient AI-PIMD simulations, these techniques will provide powerful tools to aid in the unravelling of the function of complex biological systems. 

\section{Acknowledgements}
C.M.I. and T.E.M. are supported by U.S. Department of Energy, Office of Science, Office of Basic Energy Sciences under Award Number DE-SC0014437. L.W. acknowledges a postdoctoral fellowship from the Stanford Center for Molecular Analysis and Design. T.E.M. also acknowledges support from a Cottrell Scholarship from the Research Corporation for Science Advancement and an Alfred P. Sloan Research fellowship. 

\section{References}
\bibliographystyle{elsarticle-harv}

\bibliography{EnzymeReview}

\begin{thebibliography}{132}
\expandafter\ifx\csname natexlab\endcsname\relax\def\natexlab#1{#1}\fi
\providecommand{\url}[1]{\texttt{#1}}
\providecommand{\href}[2]{#2}
\providecommand{\path}[1]{#1}
\providecommand{\DOIprefix}{doi:}
\providecommand{\ArXivprefix}{arXiv:}
\providecommand{\URLprefix}{URL: }
\providecommand{\Pubmedprefix}{pmid:}
\providecommand{\doi}[1]{\href{http://dx.doi.org/#1}{\path{#1}}}
\providecommand{\Pubmed}[1]{\href{pmid:#1}{\path{#1}}}
\providecommand{\bibinfo}[2]{#2}
\ifx\xfnm\relax \def\xfnm[#1]{\unskip,\space#1}\fi
\bibitem[{Altun et~al.(2006)Altun, Shaik and Thiel}]{Altun_2006}
\bibinfo{author}{Altun, A.}, \bibinfo{author}{Shaik, S.},
  \bibinfo{author}{Thiel, W.}, \bibinfo{year}{2006}.
\newblock \bibinfo{title}{Systematic qm/mm investigation of factors that affect
  the cytochrome p450-catalyzed hydrogen abstraction of camphor}.
\newblock \bibinfo{journal}{J. Comput. Chem.} \bibinfo{volume}{27},
  \bibinfo{pages}{1324--1337}.
\newblock \URLprefix \url{http://dx.doi.org/10.1002/jcc.20398},
  \DOIprefix\doi{10.1002/jcc.20398}.
\bibitem[{Autschbach and Srebro(2014)}]{Autschbach-2014_ACR}
\bibinfo{author}{Autschbach, J.}, \bibinfo{author}{Srebro, M.},
  \bibinfo{year}{2014}.
\newblock \bibinfo{title}{Delocalization error and “functional tuning” in
  kohn–sham calculations of molecular properties}.
\newblock \bibinfo{journal}{Accounts of Chemical Research}
  \bibinfo{volume}{47}, \bibinfo{pages}{2592--2602}.
\newblock \URLprefix \url{http://dx.doi.org/10.1021/ar500171t},
  \DOIprefix\doi{10.1021/ar500171t}.
\bibitem[{Baer et~al.({2010})Baer, Livshits and Salzner}]{Baer-2010-review}
\bibinfo{author}{Baer, R.}, \bibinfo{author}{Livshits, E.},
  \bibinfo{author}{Salzner, U.}, \bibinfo{year}{{2010}}.
\newblock \bibinfo{title}{{Tuned Range-Separated Hybrids in Density Functional
  Theory}}, in: \bibinfo{editor}{{Leone, SR and Cremer, PS and Groves, JT and
  Johnson, MA and Richmond, G}} (Ed.), \bibinfo{booktitle}{{ANNUAL REVIEW OF
  PHYSICAL CHEMISTRY, VOL 61}}. volume~\bibinfo{volume}{{61}} of
  \textit{\bibinfo{series}{{Ann. Rev. Phys. Chem.}}}, pp.
  \bibinfo{pages}{{85--109}}.
\bibitem[{Baer and Neuhauser(2005)}]{Baer-2005}
\bibinfo{author}{Baer, R.}, \bibinfo{author}{Neuhauser, D.},
  \bibinfo{year}{2005}.
\newblock \bibinfo{title}{Density functional theory with correct long-range
  asymptotic behavior}.
\newblock \bibinfo{journal}{Phys. Rev. Lett.} \bibinfo{volume}{94},
  \bibinfo{pages}{043002}.
\newblock \URLprefix
  \url{http://link.aps.org/doi/10.1103/PhysRevLett.94.043002},
  \DOIprefix\doi{10.1103/PhysRevLett.94.043002}.
\bibitem[{Becke(1993)}]{Becke1993}
\bibinfo{author}{Becke, A.D.}, \bibinfo{year}{1993}.
\newblock \bibinfo{title}{Density-functional thermochemistry. iii. the role of
  exact exchange}.
\newblock \bibinfo{journal}{J. Chem. Phys.} \bibinfo{volume}{98},
  \bibinfo{pages}{5648--5652}.
\newblock \URLprefix
  \url{http://scitation.aip.org/content/aip/journal/jcp/98/7/10.1063/1.464913},
  \DOIprefix\doi{http://dx.doi.org/10.1063/1.464913}.
\bibitem[{Becke and Johnson(2005)}]{Becke_2005}
\bibinfo{author}{Becke, A.D.}, \bibinfo{author}{Johnson, E.R.},
  \bibinfo{year}{2005}.
\newblock \bibinfo{title}{Exchange-hole dipole moment and the dispersion
  interaction}.
\newblock \bibinfo{journal}{J. Chem. Phys.} \bibinfo{volume}{122}.
\newblock \URLprefix
  \url{http://scitation.aip.org/content/aip/journal/jcp/122/15/10.1063/1.1884601},
  \DOIprefix\doi{http://dx.doi.org/10.1063/1.1884601}.
\bibitem[{Becke and Johnson(2007)}]{Becke_2007}
\bibinfo{author}{Becke, A.D.}, \bibinfo{author}{Johnson, E.R.},
  \bibinfo{year}{2007}.
\newblock \bibinfo{title}{Exchange-hole dipole moment and the dispersion
  interaction revisited}.
\newblock \bibinfo{journal}{J. Chem. Phys.} \bibinfo{volume}{127}.
\newblock \URLprefix
  \url{http://scitation.aip.org/content/aip/journal/jcp/127/15/10.1063/1.2795701},
  \DOIprefix\doi{http://dx.doi.org/10.1063/1.2795701}.
\bibitem[{Benkovic and Hammes-Schiffer(2003)}]{Benkovic2003}
\bibinfo{author}{Benkovic, S.J.}, \bibinfo{author}{Hammes-Schiffer, S.},
  \bibinfo{year}{2003}.
\newblock \bibinfo{title}{A perspective on enzyme catalysis}.
\newblock \bibinfo{journal}{Science} \bibinfo{volume}{301},
  \bibinfo{pages}{1196--1202}.
\newblock \URLprefix \url{http://science.sciencemag.org/content/301/5637/1196},
  \DOIprefix\doi{10.1126/science.1085515},
  \href{http://arxiv.org/abs/http://science.sciencemag.org/content/301/5637/1196.full.pdf}{{\tt
  arXiv:http://science.sciencemag.org/content/301/5637/1196.full.pdf}}.
\bibitem[{Berne and Thirumalai(1986)}]{bern-thir86arpc}
\bibinfo{author}{Berne, B.J.}, \bibinfo{author}{Thirumalai, D.},
  \bibinfo{year}{1986}.
\newblock \bibinfo{title}{{On the Simulation of Quantum Systems: Path Integral
  Methods}}.
\newblock \bibinfo{journal}{Ann. Rev. Phys. Chem.} \bibinfo{volume}{37},
  \bibinfo{pages}{401--424}.
\bibitem[{Bowler and Miyazaki(2012)}]{linear_scaling_2012}
\bibinfo{author}{Bowler, D.R.}, \bibinfo{author}{Miyazaki, T.},
  \bibinfo{year}{2012}.
\newblock \bibinfo{title}{O(n) methods in electronic structure calculations}.
\newblock \bibinfo{journal}{Reports on Progress in Physics}
  \bibinfo{volume}{75}, \bibinfo{pages}{036503}.
\newblock \URLprefix \url{http://stacks.iop.org/0034-4885/75/i=3/a=036503}.
\bibitem[{Bowler et~al.(2002)Bowler, Miyazaki and Gillan}]{conquest}
\bibinfo{author}{Bowler, D.R.}, \bibinfo{author}{Miyazaki, T.},
  \bibinfo{author}{Gillan, M.J.}, \bibinfo{year}{2002}.
\newblock \bibinfo{title}{Recent progress in linear scaling ab initio
  electronic structure techniques}.
\newblock \bibinfo{journal}{J. Phys.-Condens. Mat.} \bibinfo{volume}{14},
  \bibinfo{pages}{2781}.
\newblock \URLprefix \url{http://stacks.iop.org/0953-8984/14/i=11/a=303}.
\bibitem[{Burke(2012)}]{Burke_2012}
\bibinfo{author}{Burke, K.}, \bibinfo{year}{2012}.
\newblock \bibinfo{title}{Perspective on density functional theory}.
\newblock \bibinfo{journal}{J. Chem. Phys.} \bibinfo{volume}{136}.
\newblock \URLprefix
  \url{http://scitation.aip.org/content/aip/journal/jcp/136/15/10.1063/1.4704546},
  \DOIprefix\doi{http://dx.doi.org/10.1063/1.4704546}.
\bibitem[{Cao and Berne(1989)}]{cao-bern89jcp}
\bibinfo{author}{Cao, J.}, \bibinfo{author}{Berne, B.J.}, \bibinfo{year}{1989}.
\newblock \bibinfo{title}{{On energy estimators in path integral Monte Carlo
  simulations: Dependence of accuracy on algorithm}}.
\newblock \bibinfo{journal}{J. Chem. Phys.} \bibinfo{volume}{91},
  \bibinfo{pages}{6359}.
\bibitem[{Cao and Voth(1994)}]{cao-voth94jcp}
\bibinfo{author}{Cao, J.}, \bibinfo{author}{Voth, G.A.}, \bibinfo{year}{1994}.
\newblock \bibinfo{title}{{The formulation of quantum statistical mechanics
  based on the Feynman path centroid density. IV. Algorithms for centroid
  molecular dynamics}}.
\newblock \bibinfo{journal}{J. Chem. Phys.} \bibinfo{volume}{101},
  \bibinfo{pages}{6168--6183}.
\bibitem[{Case et~al.(2012)Case, Darden, Cheatham, Simmerling, Wang, Duke, Luo,
  Walker, Zhang, Merz, Roberts, Hayik, Roitberg, Seabra, Swails, Goetz,
  Kolossv\'{a}ry, Wong, Paesani, Vanicek, Wolf, Liu, Wu, Brozell, Steinbrecher,
  Gohlke, Cai, Ye, Wang, Hsieh, Cui, Roe, Mathews, Seetin, Salomon-Ferrer,
  Sagui, Babin, Luchko, Gusarov, Kovalenko and Kollman}]{amber_12}
\bibinfo{author}{Case, D.A.}, \bibinfo{author}{Darden, T.A.},
  \bibinfo{author}{Cheatham, T.E.}, \bibinfo{author}{Simmerling, C.L.},
  \bibinfo{author}{Wang, J.}, \bibinfo{author}{Duke, R.E.},
  \bibinfo{author}{Luo, R.}, \bibinfo{author}{Walker, R.C.},
  \bibinfo{author}{Zhang, W.}, \bibinfo{author}{Merz, K.M.},
  \bibinfo{author}{Roberts, B.}, \bibinfo{author}{Hayik, S.},
  \bibinfo{author}{Roitberg, A.}, \bibinfo{author}{Seabra, G.},
  \bibinfo{author}{Swails, J.}, \bibinfo{author}{Goetz, A.W.},
  \bibinfo{author}{Kolossv\'{a}ry, I.}, \bibinfo{author}{Wong, K.F.},
  \bibinfo{author}{Paesani, F.}, \bibinfo{author}{Vanicek, J.},
  \bibinfo{author}{Wolf, R.M.}, \bibinfo{author}{Liu, J.}, \bibinfo{author}{Wu,
  X.}, \bibinfo{author}{Brozell, S.R.}, \bibinfo{author}{Steinbrecher, T.},
  \bibinfo{author}{Gohlke, H.}, \bibinfo{author}{Cai, Q.}, \bibinfo{author}{Ye,
  X.}, \bibinfo{author}{Wang, J.}, \bibinfo{author}{Hsieh, M.J.},
  \bibinfo{author}{Cui, G.}, \bibinfo{author}{Roe, D.R.},
  \bibinfo{author}{Mathews, D.H.}, \bibinfo{author}{Seetin, M.G.},
  \bibinfo{author}{Salomon-Ferrer, R.}, \bibinfo{author}{Sagui, C.},
  \bibinfo{author}{Babin, V.}, \bibinfo{author}{Luchko, T.},
  \bibinfo{author}{Gusarov, S.}, \bibinfo{author}{Kovalenko, A.},
  \bibinfo{author}{Kollman, P.A.}, \bibinfo{year}{2012}.
\newblock \bibinfo{title}{Amber 12}.
\newblock \URLprefix \url{http://ambermd.org/}.
\bibitem[{Ceriotti(2010)}]{gle4md}
\bibinfo{author}{Ceriotti, M.}, \bibinfo{year}{2010}.
\newblock \bibinfo{title}{{GLE4MD}}.
\newblock \bibinfo{howpublished}{epfl-cosmo.github.io/gle4md/}.
\bibitem[{Ceriotti et~al.(2010a)Ceriotti, Bussi and Parrinello}]{ceri+10jctc}
\bibinfo{author}{Ceriotti, M.}, \bibinfo{author}{Bussi, G.},
  \bibinfo{author}{Parrinello, M.}, \bibinfo{year}{2010}a.
\newblock \bibinfo{title}{{Colored-Noise Thermostats \`{a} la Carte}}.
\newblock \bibinfo{journal}{J. Chem. Theory Comput.} \bibinfo{volume}{6},
  \bibinfo{pages}{1170--1180}.
\bibitem[{Ceriotti et~al.(2013)Ceriotti, Cuny, Parrinello and
  Manolopoulos}]{ceri+13pnas}
\bibinfo{author}{Ceriotti, M.}, \bibinfo{author}{Cuny, J.},
  \bibinfo{author}{Parrinello, M.}, \bibinfo{author}{Manolopoulos, D.E.},
  \bibinfo{year}{2013}.
\newblock \bibinfo{title}{{Nuclear quantum effects and hydrogen bond
  fluctuations in water.}}
\newblock \bibinfo{journal}{Proc. Natl. Acad. Sci. USA} \bibinfo{volume}{110},
  \bibinfo{pages}{15591--6}.
\bibitem[{Ceriotti and Manolopoulos(2012)}]{ceri-mano12prl}
\bibinfo{author}{Ceriotti, M.}, \bibinfo{author}{Manolopoulos, D.E.},
  \bibinfo{year}{2012}.
\newblock \bibinfo{title}{{Efficient First-Principles Calculation of the
  Quantum Kinetic Energy and Momentum Distribution of Nuclei}}.
\newblock \bibinfo{journal}{Phys. Rev. Lett.} \bibinfo{volume}{109},
  \bibinfo{pages}{100604}.
\bibitem[{Ceriotti et~al.(2011)Ceriotti, Manolopoulos and
  Parrinello}]{ceri+11jcp}
\bibinfo{author}{Ceriotti, M.}, \bibinfo{author}{Manolopoulos, D.E.},
  \bibinfo{author}{Parrinello, M.}, \bibinfo{year}{2011}.
\newblock \bibinfo{title}{{Accelerating the convergence of path integral
  dynamics with a generalized Langevin equation.}}
\newblock \bibinfo{journal}{J. Chem. Phys.} \bibinfo{volume}{134},
  \bibinfo{pages}{84104}.
\bibitem[{Ceriotti and Markland(2013)}]{ceri-mark13jcp}
\bibinfo{author}{Ceriotti, M.}, \bibinfo{author}{Markland, T.E.},
  \bibinfo{year}{2013}.
\newblock \bibinfo{title}{{Efficient methods and practical guidelines for
  simulating isotope effects.}}
\newblock \bibinfo{journal}{J. Chem. Phys.} \bibinfo{volume}{138},
  \bibinfo{pages}{014112}.
\bibitem[{Ceriotti et~al.(2010b)Ceriotti, Parrinello, Markland and
  Manolopoulos}]{ceri+10jcp}
\bibinfo{author}{Ceriotti, M.}, \bibinfo{author}{Parrinello, M.},
  \bibinfo{author}{Markland, T.E.}, \bibinfo{author}{Manolopoulos, D.E.},
  \bibinfo{year}{2010}b.
\newblock \bibinfo{title}{{Efficient stochastic thermostatting of path integral
  molecular dynamics.}}
\newblock \bibinfo{journal}{J. Chem. Phys.} \bibinfo{volume}{133},
  \bibinfo{pages}{124104}.
\bibitem[{Chai and Head-Gordon({2008})}]{Head-Gordon-2008}
\bibinfo{author}{Chai, J.D.}, \bibinfo{author}{Head-Gordon, M.},
  \bibinfo{year}{{2008}}.
\newblock \bibinfo{title}{{Systematic optimization of long-range corrected
  hybrid density functionals}}.
\newblock \bibinfo{journal}{{J. Chem. Phys.}} \bibinfo{volume}{{128}},
  \bibinfo{pages}{{084106}}.
\bibitem[{Chang et~al.(2004)Chang, Chang, Damrauer and Nocera}]{Chang2004}
\bibinfo{author}{Chang, C.J.}, \bibinfo{author}{Chang, M.C.},
  \bibinfo{author}{Damrauer, N.H.}, \bibinfo{author}{Nocera, D.G.},
  \bibinfo{year}{2004}.
\newblock \bibinfo{title}{Proton-coupled electron transfer: a unifying
  mechanism for biological charge transport, amino acid radical initiation and
  propagation, and bond making/breaking reactions of water and oxygen}.
\newblock \bibinfo{journal}{BBA-Bioenergetics} \bibinfo{volume}{1655},
  \bibinfo{pages}{13 -- 28}.
\bibitem[{Cleland et~al.(1998)Cleland, Frey and Gerlt}]{Cleland1998}
\bibinfo{author}{Cleland, W.W.}, \bibinfo{author}{Frey, P.A.},
  \bibinfo{author}{Gerlt, J.A.}, \bibinfo{year}{1998}.
\newblock \bibinfo{title}{The low barrier hydrogen bond in enzymatic
  catalysis}.
\newblock \bibinfo{journal}{J. Biol. Chem.} \bibinfo{volume}{273},
  \bibinfo{pages}{25529--25532}.
\newblock \URLprefix \url{http://www.jbc.org/content/273/40/25529.short}.
\bibitem[{Cohen et~al.(2008a)Cohen, Mori-S\'{a}nchez and Yang}]{Weitao-2008-d}
\bibinfo{author}{Cohen, A.J.}, \bibinfo{author}{Mori-S\'{a}nchez, P.},
  \bibinfo{author}{Yang, W.}, \bibinfo{year}{2008}a.
\newblock \bibinfo{title}{Fractional charge perspective on the band gap in
  density-functional theory}.
\newblock \bibinfo{journal}{Phys. Rev. B} \bibinfo{volume}{77},
  \bibinfo{pages}{115123}.
\bibitem[{Cohen et~al.(2008b)Cohen, Mori-S\'{a}nchez and Yang}]{Weitao-2008-c}
\bibinfo{author}{Cohen, A.J.}, \bibinfo{author}{Mori-S\'{a}nchez, P.},
  \bibinfo{author}{Yang, W.}, \bibinfo{year}{2008}b.
\newblock \bibinfo{title}{Insights into current limitations of density
  functional theory}.
\newblock \bibinfo{journal}{Science} \bibinfo{volume}{321},
  \bibinfo{pages}{792--794}.
\bibitem[{Craig and Manolopoulos(2004)}]{crai-mano04jcp}
\bibinfo{author}{Craig, I.R.}, \bibinfo{author}{Manolopoulos, D.E.},
  \bibinfo{year}{2004}.
\newblock \bibinfo{title}{{Quantum statistics and classical mechanics: Real
  time correlation functions from ring polymer molecular dynamics}}.
\newblock \bibinfo{journal}{J. Chem. Phys.} \bibinfo{volume}{121},
  \bibinfo{pages}{3368}.
\bibitem[{Cukier(2004)}]{Cukier2004}
\bibinfo{author}{Cukier, R.}, \bibinfo{year}{2004}.
\newblock \bibinfo{title}{Theory and simulation of proton-coupled electron
  transfer, hydrogen-atom transfer, and proton translocation in proteins}.
\newblock \bibinfo{journal}{BBA-Bioenergetics} \bibinfo{volume}{1655},
  \bibinfo{pages}{37 -- 44}.
\bibitem[{Duan et~al.(2003)Duan, Wu, Chowdhury, Lee, Xiong, Zhang, Yang,
  Cieplak, Luo, Lee, Caldwell, Wang and Kollman}]{JCC_2003}
\bibinfo{author}{Duan, Y.}, \bibinfo{author}{Wu, C.},
  \bibinfo{author}{Chowdhury, S.}, \bibinfo{author}{Lee, M.C.},
  \bibinfo{author}{Xiong, G.}, \bibinfo{author}{Zhang, W.},
  \bibinfo{author}{Yang, R.}, \bibinfo{author}{Cieplak, P.},
  \bibinfo{author}{Luo, R.}, \bibinfo{author}{Lee, T.},
  \bibinfo{author}{Caldwell, J.}, \bibinfo{author}{Wang, J.},
  \bibinfo{author}{Kollman, P.}, \bibinfo{year}{2003}.
\newblock \bibinfo{title}{A point-charge force field for molecular mechanics
  simulations of proteins based on condensed-phase quantum mechanical
  calculations}.
\newblock \bibinfo{journal}{J. Comput. Chem.} \bibinfo{volume}{24},
  \bibinfo{pages}{1999--2012}.
\newblock \URLprefix \url{http://dx.doi.org/10.1002/jcc.10349},
  \DOIprefix\doi{10.1002/jcc.10349}.
\bibitem[{Elstner et~al.(2001)Elstner, Hobza, Frauenheim, Suhai and
  Kaxiras}]{Elstner_2001}
\bibinfo{author}{Elstner, M.}, \bibinfo{author}{Hobza, P.},
  \bibinfo{author}{Frauenheim, T.}, \bibinfo{author}{Suhai, S.},
  \bibinfo{author}{Kaxiras, E.}, \bibinfo{year}{2001}.
\newblock \bibinfo{title}{Hydrogen bonding and stacking interactions of nucleic
  acid base pairs: A density-functional-theory based treatment}.
\newblock \bibinfo{journal}{J. Chem. Phys.} \bibinfo{volume}{114},
  \bibinfo{pages}{5149--5155}.
\newblock \URLprefix
  \url{http://scitation.aip.org/content/aip/journal/jcp/114/12/10.1063/1.1329889},
  \DOIprefix\doi{http://dx.doi.org/10.1063/1.1329889}.
\bibitem[{Fanourgakis et~al.(2009)Fanourgakis, Markland and
  Manolopoulos}]{Fanourgakis2009}
\bibinfo{author}{Fanourgakis, G.S.}, \bibinfo{author}{Markland, T.E.},
  \bibinfo{author}{Manolopoulos, D.E.}, \bibinfo{year}{2009}.
\newblock \bibinfo{title}{A fast path integral method for polarizable force
  fields}.
\newblock \bibinfo{journal}{J. Chem. Phys.} \bibinfo{volume}{131}.
\newblock \URLprefix
  \url{http://scitation.aip.org/content/aip/journal/jcp/131/9/10.1063/1.3216520},
  \DOIprefix\doi{http://dx.doi.org/10.1063/1.3216520}.
\bibitem[{Feynman and Hibbs(1964)}]{feyn-hibb65book}
\bibinfo{author}{Feynman, R.P.}, \bibinfo{author}{Hibbs, A.R.},
  \bibinfo{year}{1964}.
\newblock \bibinfo{title}{{Quantum Mechanics and Path Integrals}}.
\newblock \bibinfo{publisher}{McGraw-Hill}, \bibinfo{address}{New York}.
\bibitem[{Flaig et~al.(2012)Flaig, Beer and Ochsenfeld}]{Flaig_2012}
\bibinfo{author}{Flaig, D.}, \bibinfo{author}{Beer, M.},
  \bibinfo{author}{Ochsenfeld, C.}, \bibinfo{year}{2012}.
\newblock \bibinfo{title}{Convergence of electronic structure with the size of
  the qm region: Example of qm/mm nmr shieldings}.
\newblock \bibinfo{journal}{J. Chem.Theory Comput.} \bibinfo{volume}{8},
  \bibinfo{pages}{2260--2271}.
\newblock \URLprefix \url{http://dx.doi.org/10.1021/ct300036s},
  \DOIprefix\doi{10.1021/ct300036s},
  \href{http://arxiv.org/abs/http://dx.doi.org/10.1021/ct300036s}{{\tt
  arXiv:http://dx.doi.org/10.1021/ct300036s}}.
\bibitem[{Fox et~al.(2011)Fox, Pittock, Fox, Tautermann, Malcolm and
  Skylaris}]{Fox_2011}
\bibinfo{author}{Fox, S.J.}, \bibinfo{author}{Pittock, C.},
  \bibinfo{author}{Fox, T.}, \bibinfo{author}{Tautermann, C.S.},
  \bibinfo{author}{Malcolm, N.}, \bibinfo{author}{Skylaris, C.K.},
  \bibinfo{year}{2011}.
\newblock \bibinfo{title}{Electrostatic embedding in large-scale first
  principles quantum mechanical calculations on biomolecules}.
\newblock \bibinfo{journal}{J. Chem. Phys.} \bibinfo{volume}{135}.
\newblock \URLprefix
  \url{http://scitation.aip.org/content/aip/journal/jcp/135/22/10.1063/1.3665893},
  \DOIprefix\doi{http://dx.doi.org/10.1063/1.3665893}.
\bibitem[{Friesner and Guallar(2005)}]{Friesner_2005}
\bibinfo{author}{Friesner, R.A.}, \bibinfo{author}{Guallar, V.},
  \bibinfo{year}{2005}.
\newblock \bibinfo{title}{Ab initio quantum chemical and mixed quantum
  mechanics/molecular mechanics (qm/mm) methods for studying enzymatic
  catalysis}.
\newblock \bibinfo{journal}{Annual Review of Physical Chemistry}
  \bibinfo{volume}{56}, \bibinfo{pages}{389--427}.
\newblock \URLprefix
  \url{http://dx.doi.org/10.1146/annurev.physchem.55.091602.094410},
  \DOIprefix\doi{10.1146/annurev.physchem.55.091602.094410},
  \href{http://arxiv.org/abs/http://dx.doi.org/10.1146/annurev.physchem.55.091602.094410}{{\tt
  arXiv:http://dx.doi.org/10.1146/annurev.physchem.55.091602.094410}}.
\bibitem[{Gao and Truhlar(2002)}]{Gao2002}
\bibinfo{author}{Gao, J.}, \bibinfo{author}{Truhlar, D.G.},
  \bibinfo{year}{2002}.
\newblock \bibinfo{title}{Quantum mechanical methods for enzyme kinetics}.
\newblock \bibinfo{journal}{Annu. Rev. Phys. Chem.} \bibinfo{volume}{53},
  \bibinfo{pages}{467--505}.
\newblock \URLprefix
  \url{http://dx.doi.org/10.1146/annurev.physchem.53.091301.150114},
  \DOIprefix\doi{10.1146/annurev.physchem.53.091301.150114}.
\bibitem[{Gaus et~al.(2014)Gaus, Cui and Elstner}]{Gaus2014}
\bibinfo{author}{Gaus, M.}, \bibinfo{author}{Cui, Q.},
  \bibinfo{author}{Elstner, M.}, \bibinfo{year}{2014}.
\newblock \bibinfo{title}{Density functional tight binding: application to
  organic and biological molecules}.
\newblock \bibinfo{journal}{Wiley Interdisciplinary Reviews: Computational
  Molecular Science} \bibinfo{volume}{4}, \bibinfo{pages}{49--61}.
\newblock \URLprefix \url{http://dx.doi.org/10.1002/wcms.1156},
  \DOIprefix\doi{10.1002/wcms.1156}.
\bibitem[{Gill et~al.(1996)Gill, Adamson and Pople}]{gill1996}
\bibinfo{author}{Gill, P.M.W.}, \bibinfo{author}{Adamson, R.D.},
  \bibinfo{author}{Pople, J.A.}, \bibinfo{year}{1996}.
\newblock \bibinfo{title}{Coulomb-attenuated exchange energy density
  functionals}.
\newblock \bibinfo{journal}{Mol. Phys.} \bibinfo{volume}{88},
  \bibinfo{pages}{1005--1009}.
\bibitem[{Goedecker(1999)}]{linear_scaling_1999}
\bibinfo{author}{Goedecker, S.}, \bibinfo{year}{1999}.
\newblock \bibinfo{title}{Linear scaling electronic structure methods}.
\newblock \bibinfo{journal}{Rev. Mod. Phys.} \bibinfo{volume}{71},
  \bibinfo{pages}{1085--1123}.
\newblock \URLprefix \url{http://link.aps.org/doi/10.1103/RevModPhys.71.1085},
  \DOIprefix\doi{10.1103/RevModPhys.71.1085}.
\bibitem[{Grimme(2006)}]{Grimme_2006}
\bibinfo{author}{Grimme, S.}, \bibinfo{year}{2006}.
\newblock \bibinfo{title}{Semiempirical gga-type density functional constructed
  with a long-range dispersion correction}.
\newblock \bibinfo{journal}{J. Comput. Chem.} \bibinfo{volume}{27},
  \bibinfo{pages}{1787--1799}.
\newblock \URLprefix \url{http://dx.doi.org/10.1002/jcc.20495},
  \DOIprefix\doi{10.1002/jcc.20495}.
\bibitem[{Grimme(2011)}]{Grimme_2011}
\bibinfo{author}{Grimme, S.}, \bibinfo{year}{2011}.
\newblock \bibinfo{title}{Density functional theory with london dispersion
  corrections}.
\newblock \bibinfo{journal}{Wiley Interdisciplinary Reviews: Computational
  Molecular Science} \bibinfo{volume}{1}, \bibinfo{pages}{211--228}.
\newblock \URLprefix \url{http://dx.doi.org/10.1002/wcms.30},
  \DOIprefix\doi{10.1002/wcms.30}.
\bibitem[{Grimme et~al.(2010)Grimme, Antony, Ehrlich and Krieg}]{grim+10jcp}
\bibinfo{author}{Grimme, S.}, \bibinfo{author}{Antony, J.},
  \bibinfo{author}{Ehrlich, S.}, \bibinfo{author}{Krieg, H.},
  \bibinfo{year}{2010}.
\newblock \bibinfo{title}{{A consistent and accurate ab initio parametrization
  of density functional dispersion correction (DFT-D) for the 94 elements
  H-Pu.}}
\newblock \bibinfo{journal}{J. Chem. Phys.} \bibinfo{volume}{132},
  \bibinfo{pages}{154104}.
\bibitem[{Ha et~al.(2000)Ha, Kim, Lee, Choi and Oh}]{Ha2000}
\bibinfo{author}{Ha, N.C.}, \bibinfo{author}{Kim, M.S.}, \bibinfo{author}{Lee,
  W.}, \bibinfo{author}{Choi, K.Y.}, \bibinfo{author}{Oh, B.H.},
  \bibinfo{year}{2000}.
\newblock \bibinfo{title}{Detection of large pka perturbations of an inhibitor
  and a catalytic group at an enzyme active site, a mechanistic basis for
  catalytic power of many enzymes}.
\newblock \bibinfo{journal}{J. Biol. Chem.} \bibinfo{volume}{275},
  \bibinfo{pages}{41100--41106}.
\newblock \URLprefix \url{http://www.jbc.org/content/275/52/41100.abstract}.
\bibitem[{Habershon et~al.(2013)Habershon, Manolopoulos, Markland and
  III}]{Habershon2013}
\bibinfo{author}{Habershon, S.}, \bibinfo{author}{Manolopoulos, D.E.},
  \bibinfo{author}{Markland, T.E.}, \bibinfo{author}{III, T.F.M.},
  \bibinfo{year}{2013}.
\newblock \bibinfo{title}{Ring-polymer molecular dynamics: Quantum effects in
  chemical dynamics from classical trajectories in an extended phase space}.
\newblock \bibinfo{journal}{Annual Review of Physical Chemistry}
  \bibinfo{volume}{64}, \bibinfo{pages}{387--413}.
\newblock \URLprefix
  \url{http://dx.doi.org/10.1146/annurev-physchem-040412-110122},
  \DOIprefix\doi{10.1146/annurev-physchem-040412-110122},
  \href{http://arxiv.org/abs/http://dx.doi.org/10.1146/annurev-physchem-040412-110122}{{\tt
  arXiv:http://dx.doi.org/10.1146/annurev-physchem-040412-110122}}.
\bibitem[{Hammes-Schiffer and Stuchebrukhov(2010)}]{Hammes-Schiffer2010}
\bibinfo{author}{Hammes-Schiffer, S.}, \bibinfo{author}{Stuchebrukhov, A.A.},
  \bibinfo{year}{2010}.
\newblock \bibinfo{title}{Theory of coupled electron and proton transfer
  reactions}.
\newblock \bibinfo{journal}{Chem. Rev.} \bibinfo{volume}{110},
  \bibinfo{pages}{6939--6960}.
\newblock \DOIprefix\doi{10.1021/cr1001436},
  \href{http://arxiv.org/abs/http://dx.doi.org/10.1021/cr1001436}{{\tt
  arXiv:http://dx.doi.org/10.1021/cr1001436}}.
\bibitem[{Haworth and Bacskay(2002)}]{Haworth_2002}
\bibinfo{author}{Haworth, N.L.}, \bibinfo{author}{Bacskay, G.B.},
  \bibinfo{year}{2002}.
\newblock \bibinfo{title}{Heats of formation of phosphorus compounds determined
  by current methods of computational quantum chemistry}.
\newblock \bibinfo{journal}{J. Chem. Phys.} \bibinfo{volume}{117},
  \bibinfo{pages}{11175--11187}.
\newblock \URLprefix
  \url{http://scitation.aip.org/content/aip/journal/jcp/117/24/10.1063/1.1521760},
  \DOIprefix\doi{http://dx.doi.org/10.1063/1.1521760}.
\bibitem[{Heaton-Burgess and Yang({2010})}]{Weitao-2010}
\bibinfo{author}{Heaton-Burgess, T.}, \bibinfo{author}{Yang, W.},
  \bibinfo{year}{{2010}}.
\newblock \bibinfo{title}{{Structural manifestation of the delocalization error
  of density functional approximations: C4N+2 rings and C-20 bowl, cage, and
  ring isomers}}.
\newblock \bibinfo{journal}{{J. Chem. Phys.}} \bibinfo{volume}{{132}},
  \bibinfo{pages}{{234113}}.
\bibitem[{Henry et~al.(2002)Henry, Parkinson and Radom}]{Henry_2002}
\bibinfo{author}{Henry, D.J.}, \bibinfo{author}{Parkinson, C.J.},
  \bibinfo{author}{Radom, L.}, \bibinfo{year}{2002}.
\newblock \bibinfo{title}{An assessment of the performance of high-level
  theoretical procedures in the computation of the heats of formation of small
  open-shell molecules}.
\newblock \bibinfo{journal}{J. Phys. Chem. A} \bibinfo{volume}{106},
  \bibinfo{pages}{7927--7936}.
\newblock \URLprefix \url{http://dx.doi.org/10.1021/jp0260752},
  \DOIprefix\doi{10.1021/jp0260752},
  \href{http://arxiv.org/abs/http://dx.doi.org/10.1021/jp0260752}{{\tt
  arXiv:http://dx.doi.org/10.1021/jp0260752}}.
\bibitem[{Herman et~al.(1982)Herman, Bruskin and Berne}]{herm-bern82jcp}
\bibinfo{author}{Herman, M.F.}, \bibinfo{author}{Bruskin, E.J.},
  \bibinfo{author}{Berne, B.J.}, \bibinfo{year}{1982}.
\newblock \bibinfo{title}{{On path integral Monte Carlo simulations}}.
\newblock \bibinfo{journal}{J. Chem. Phys.} \bibinfo{volume}{76},
  \bibinfo{pages}{5150}.
\bibitem[{Hohenberg and Kohn(1964)}]{Hohenberg_Kohn}
\bibinfo{author}{Hohenberg, P.}, \bibinfo{author}{Kohn, W.},
  \bibinfo{year}{1964}.
\newblock \bibinfo{title}{Inhomogeneous electron gas}.
\newblock \bibinfo{journal}{Phys. Rev.} \bibinfo{volume}{136},
  \bibinfo{pages}{B864--B871}.
\newblock \URLprefix \url{http://link.aps.org/doi/10.1103/PhysRev.136.B864},
  \DOIprefix\doi{10.1103/PhysRev.136.B864}.
\bibitem[{Hu et~al.(2011)Hu, Söderhjelm and Ryde}]{Hu_2011}
\bibinfo{author}{Hu, L.}, \bibinfo{author}{Söderhjelm, P.},
  \bibinfo{author}{Ryde, U.}, \bibinfo{year}{2011}.
\newblock \bibinfo{title}{On the convergence of qm/mm energies}.
\newblock \bibinfo{journal}{J. Chem. Theory Comput.} \bibinfo{volume}{7},
  \bibinfo{pages}{761--777}.
\newblock \URLprefix \url{http://dx.doi.org/10.1021/ct100530r},
  \DOIprefix\doi{10.1021/ct100530r},
  \href{http://arxiv.org/abs/http://dx.doi.org/10.1021/ct100530r}{{\tt
  arXiv:http://dx.doi.org/10.1021/ct100530r}}.
\bibitem[{Iikura et~al.(2001)Iikura, Tsuneda, Yanai and Hirao}]{iikura2001}
\bibinfo{author}{Iikura, H.}, \bibinfo{author}{Tsuneda, T.},
  \bibinfo{author}{Yanai, T.}, \bibinfo{author}{Hirao, K.},
  \bibinfo{year}{2001}.
\newblock \bibinfo{title}{A long-range correction scheme for
  generalized-gradient-approximation exchange functionals}.
\newblock \bibinfo{journal}{J. Chem. Phys.} \bibinfo{volume}{115},
  \bibinfo{pages}{3540--3544}.
\bibitem[{Isborn et~al.(2012)Isborn, Götz, Clark, Walker and
  Martínez}]{Isborn_2012}
\bibinfo{author}{Isborn, C.M.}, \bibinfo{author}{Götz, A.W.},
  \bibinfo{author}{Clark, M.A.}, \bibinfo{author}{Walker, R.C.},
  \bibinfo{author}{Martínez, T.J.}, \bibinfo{year}{2012}.
\newblock \bibinfo{title}{Electronic absorption spectra from mm and ab initio
  qm/mm molecular dynamics: Environmental effects on the absorption spectrum of
  photoactive yellow protein}.
\newblock \bibinfo{journal}{J. Chem. Theory Comput.} \bibinfo{volume}{8},
  \bibinfo{pages}{5092--5106}.
\newblock \URLprefix \url{http://dx.doi.org/10.1021/ct3006826},
  \DOIprefix\doi{10.1021/ct3006826},
  \href{http://arxiv.org/abs/http://dx.doi.org/10.1021/ct3006826}{{\tt
  arXiv:http://dx.doi.org/10.1021/ct3006826}}.
\bibitem[{Isborn et~al.(2011)Isborn, Luehr, Ufimtsev and
  Martínez}]{Isborn_2011}
\bibinfo{author}{Isborn, C.M.}, \bibinfo{author}{Luehr, N.},
  \bibinfo{author}{Ufimtsev, I.S.}, \bibinfo{author}{Martínez, T.J.},
  \bibinfo{year}{2011}.
\newblock \bibinfo{title}{Excited-state electronic structure with configuration
  interaction singles and tamm-dancoff time-dependent density functional theory
  on graphical processing units}.
\newblock \bibinfo{journal}{J. Chem. Theory Comput.} \bibinfo{volume}{7},
  \bibinfo{pages}{1814--1823}.
\newblock \URLprefix \url{http://dx.doi.org/10.1021/ct200030k},
  \DOIprefix\doi{10.1021/ct200030k},
  \href{http://arxiv.org/abs/http://dx.doi.org/10.1021/ct200030k}{{\tt
  arXiv:http://dx.doi.org/10.1021/ct200030k}}.
\bibitem[{Isborn et~al.(2013)Isborn, Mar, Curchod, Tavernelli and
  Martínez}]{Isborn_2013}
\bibinfo{author}{Isborn, C.M.}, \bibinfo{author}{Mar, B.D.},
  \bibinfo{author}{Curchod, B.F.E.}, \bibinfo{author}{Tavernelli, I.},
  \bibinfo{author}{Martínez, T.J.}, \bibinfo{year}{2013}.
\newblock \bibinfo{title}{The charge transfer problem in density functional
  theory calculations of aqueously solvated molecules}.
\newblock \bibinfo{journal}{J. Phys. Chem. B} \bibinfo{volume}{117},
  \bibinfo{pages}{12189--12201}.
\newblock \URLprefix \url{http://dx.doi.org/10.1021/jp4058274},
  \DOIprefix\doi{10.1021/jp4058274},
  \href{http://arxiv.org/abs/http://dx.doi.org/10.1021/jp4058274}{{\tt
  arXiv:http://dx.doi.org/10.1021/jp4058274}}.
\bibitem[{Jang and Voth(1999)}]{Jang1999}
\bibinfo{author}{Jang, S.}, \bibinfo{author}{Voth, G.A.}, \bibinfo{year}{1999}.
\newblock \bibinfo{title}{A derivation of centroid molecular dynamics and other
  approximate time evolution methods for path integral centroid variables}.
\newblock \bibinfo{journal}{J. Chem. Phys.} \bibinfo{volume}{111},
  \bibinfo{pages}{2371--2384}.
\newblock \URLprefix
  \url{http://scitation.aip.org/content/aip/journal/jcp/111/6/10.1063/1.479515},
  \DOIprefix\doi{http://dx.doi.org/10.1063/1.479515}.
\bibitem[{Johnson and Becke(2006)}]{Johnson_2006}
\bibinfo{author}{Johnson, E.R.}, \bibinfo{author}{Becke, A.D.},
  \bibinfo{year}{2006}.
\newblock \bibinfo{title}{A post-hartree-fock model of intermolecular
  interactions: Inclusion of higher-order corrections}.
\newblock \bibinfo{journal}{J. Chem. Phys.} \bibinfo{volume}{124}.
\newblock \URLprefix
  \url{http://scitation.aip.org/content/aip/journal/jcp/124/17/10.1063/1.2190220},
  \DOIprefix\doi{http://dx.doi.org/10.1063/1.2190220}.
\bibitem[{Johnson et~al.(2009)Johnson, Mackie and DiLabio}]{Johnson_2009}
\bibinfo{author}{Johnson, E.R.}, \bibinfo{author}{Mackie, I.D.},
  \bibinfo{author}{DiLabio, G.A.}, \bibinfo{year}{2009}.
\newblock \bibinfo{title}{Dispersion interactions in density-functional
  theory}.
\newblock \bibinfo{journal}{J. Phys. Org. Chem.} \bibinfo{volume}{22},
  \bibinfo{pages}{1127--1135}.
\newblock \URLprefix \url{http://dx.doi.org/10.1002/poc.1606},
  \DOIprefix\doi{10.1002/poc.1606}.
\bibitem[{Jorgensen et~al.(1983)Jorgensen, Chandrasekhar, Madura, Impey and
  Klein}]{jorg+83jcp}
\bibinfo{author}{Jorgensen, W.W.L.}, \bibinfo{author}{Chandrasekhar, J.},
  \bibinfo{author}{Madura, J.D.}, \bibinfo{author}{Impey, R.W.},
  \bibinfo{author}{Klein, M.L.}, \bibinfo{year}{1983}.
\newblock \bibinfo{title}{{Comparison of simple potential functions for
  simulating liquid water}}.
\newblock \bibinfo{journal}{J. Chem. Phys.} \bibinfo{volume}{79},
  \bibinfo{pages}{926}.
\bibitem[{Kapil et~al.(2016)Kapil, VandeVondele and Ceriotti}]{Kapil2016}
\bibinfo{author}{Kapil, V.}, \bibinfo{author}{VandeVondele, J.},
  \bibinfo{author}{Ceriotti, M.}, \bibinfo{year}{2016}.
\newblock \bibinfo{title}{Accurate molecular dynamics and nuclear quantum
  effects at low cost by multiple steps in real and imaginary time: Using
  density functional theory to accelerate wavefunction methods}.
\newblock \bibinfo{journal}{J. Chem. Phys.} \bibinfo{volume}{144},
  \bibinfo{pages}{054111}.
\bibitem[{Kim et~al.(2013)Kim, Sim and Burke}]{kim2013}
\bibinfo{author}{Kim, M.C.}, \bibinfo{author}{Sim, E.}, \bibinfo{author}{Burke,
  K.}, \bibinfo{year}{2013}.
\newblock \bibinfo{title}{Understanding and reducing errors in density
  functional calculations}.
\newblock \bibinfo{journal}{Phys. Rev. Lett.} \bibinfo{volume}{111},
  \bibinfo{pages}{073003}.
\bibitem[{Klinman and Kohen(2013)}]{Klinman2013}
\bibinfo{author}{Klinman, J.P.}, \bibinfo{author}{Kohen, A.},
  \bibinfo{year}{2013}.
\newblock \bibinfo{title}{Hydrogen tunneling links protein dynamics to enzyme
  catalysis}.
\newblock \bibinfo{journal}{Annu. Rev. Biochem.} \bibinfo{volume}{82},
  \bibinfo{pages}{471--496}.
\newblock \URLprefix
  \url{http://dx.doi.org/10.1146/annurev-biochem-051710-133623},
  \DOIprefix\doi{10.1146/annurev-biochem-051710-133623}.
\bibitem[{Kohen and H(2006)}]{Kohen2005}
\bibinfo{editor}{Kohen, A.}, \bibinfo{editor}{H, L.} (Eds.),
  \bibinfo{year}{2006}.
\newblock \bibinfo{title}{{Isotope effects in Chemistry and Biology}}.
\newblock \bibinfo{publisher}{CRC Press}, \bibinfo{address}{Florida}.
\bibitem[{Kohn and Sham(1965)}]{Kohn_Sham_1965}
\bibinfo{author}{Kohn, W.}, \bibinfo{author}{Sham, L.J.}, \bibinfo{year}{1965}.
\newblock \bibinfo{title}{Self-consistent equations including exchange and
  correlation effects}.
\newblock \bibinfo{journal}{Phys. Rev.} \bibinfo{volume}{140},
  \bibinfo{pages}{A1133--A1138}.
\newblock \URLprefix \url{http://link.aps.org/doi/10.1103/PhysRev.140.A1133},
  \DOIprefix\doi{10.1103/PhysRev.140.A1133}.
\bibitem[{Kraisler and Kronik({2014})}]{Kronik-2014}
\bibinfo{author}{Kraisler, E.}, \bibinfo{author}{Kronik, L.},
  \bibinfo{year}{{2014}}.
\newblock \bibinfo{title}{{Fundamental gaps with approximate density
  functionals: The derivative discontinuity revealed from ensemble
  considerations}}.
\newblock \bibinfo{journal}{{J. Chem. Phys.}} \bibinfo{volume}{{140}},
  \bibinfo{pages}{{18A540}}.
\bibitem[{Kronik et~al.(2012)Kronik, Stein, Refaely-Abramson and
  Baer}]{Kronik_2012}
\bibinfo{author}{Kronik, L.}, \bibinfo{author}{Stein, T.},
  \bibinfo{author}{Refaely-Abramson, S.}, \bibinfo{author}{Baer, R.},
  \bibinfo{year}{2012}.
\newblock \bibinfo{title}{Excitation gaps of finite-sized systems from
  optimally tuned range-separated hybrid functionals}.
\newblock \bibinfo{journal}{J. Chem. Theory Comput.} \bibinfo{volume}{8},
  \bibinfo{pages}{1515--1531}.
\newblock \URLprefix \url{http://dx.doi.org/10.1021/ct2009363},
  \DOIprefix\doi{10.1021/ct2009363},
  \href{http://arxiv.org/abs/http://dx.doi.org/10.1021/ct2009363}{{\tt
  arXiv:http://dx.doi.org/10.1021/ct2009363}}.
\bibitem[{Kulik et~al.(2012)Kulik, Luehr, Ufimtsev and Martinez}]{Kulik_2012}
\bibinfo{author}{Kulik, H.J.}, \bibinfo{author}{Luehr, N.},
  \bibinfo{author}{Ufimtsev, I.S.}, \bibinfo{author}{Martinez, T.J.},
  \bibinfo{year}{2012}.
\newblock \bibinfo{title}{Ab initio quantum chemistry for protein structures}.
\newblock \bibinfo{journal}{J. Phys. Chem. B} \bibinfo{volume}{116},
  \bibinfo{pages}{12501--12509}.
\newblock \URLprefix \url{http://dx.doi.org/10.1021/jp307741u},
  \DOIprefix\doi{10.1021/jp307741u},
  \href{http://arxiv.org/abs/http://dx.doi.org/10.1021/jp307741u}{{\tt
  arXiv:http://dx.doi.org/10.1021/jp307741u}}.
\bibitem[{Kuritz et~al.({2011})Kuritz, Stein, Baer and Kronik}]{Baer-2011-c}
\bibinfo{author}{Kuritz, N.}, \bibinfo{author}{Stein, T.},
  \bibinfo{author}{Baer, R.}, \bibinfo{author}{Kronik, L.},
  \bibinfo{year}{{2011}}.
\newblock \bibinfo{title}{{Charge-Transfer-Like pi(->)pi{*} Excitations in
  Time-Dependent Density Functional Theory: A Conundrum and Its Solution}}.
\newblock \bibinfo{journal}{{J. Chem. Theory Comput.}} \bibinfo{volume}{{7}},
  \bibinfo{pages}{{2408--2415}}.
\bibitem[{Lennartz et~al.(2002)Lennartz, Schäfer, Terstegen and
  Thiel}]{Lennartz_2002}
\bibinfo{author}{Lennartz, C.}, \bibinfo{author}{Schäfer, A.},
  \bibinfo{author}{Terstegen, F.}, \bibinfo{author}{Thiel, W.},
  \bibinfo{year}{2002}.
\newblock \bibinfo{title}{Enzymatic reactions of triosephosphate isomerase: A
  theoretical calibration study}.
\newblock \bibinfo{journal}{J. Phys. Chem. B} \bibinfo{volume}{106},
  \bibinfo{pages}{1758--1767}.
\newblock \URLprefix \url{http://dx.doi.org/10.1021/jp012658k},
  \DOIprefix\doi{10.1021/jp012658k},
  \href{http://arxiv.org/abs/http://dx.doi.org/10.1021/jp012658k}{{\tt
  arXiv:http://dx.doi.org/10.1021/jp012658k}}.
\bibitem[{Lin and Truhlar(2006)}]{Lin2006}
\bibinfo{author}{Lin, H.}, \bibinfo{author}{Truhlar, D.G.},
  \bibinfo{year}{2006}.
\newblock \bibinfo{title}{Qm/mm: what have we learned, where are we, and where
  do we go from here?}
\newblock \bibinfo{journal}{Theoretical Chemistry Accounts}
  \bibinfo{volume}{117}, \bibinfo{pages}{185--199}.
\newblock \URLprefix \url{http://dx.doi.org/10.1007/s00214-006-0143-z},
  \DOIprefix\doi{10.1007/s00214-006-0143-z}.
\bibitem[{Luehr et~al.(2016)Luehr, Sisto and Martinez}]{Luehr2016}
\bibinfo{author}{Luehr, N.}, \bibinfo{author}{Sisto, A.},
  \bibinfo{author}{Martinez, T.J.}, \bibinfo{year}{2016}.
\newblock \bibinfo{title}{Gaussian basis set hartree-fock, density functional
  theory, and beyond on gpus}, in: \bibinfo{editor}{Walker, R.},
  \bibinfo{editor}{Goetz, A.} (Eds.), \bibinfo{booktitle}{Electronic Structure
  Calculations on Graphics Processing Units: From Quantum Chemistry to
  Condensed Matter Physics}. \bibinfo{publisher}{Wiley}, \bibinfo{address}{West
  Sussex}. chapter~\bibinfo{chapter}{4}, pp. \bibinfo{pages}{67--100}.
\bibitem[{López-Canut et~al.(2009)López-Canut, Martí, Bertrán, Moliner and
  Tuñón}]{Lopez-Canut_2009}
\bibinfo{author}{López-Canut, V.}, \bibinfo{author}{Martí, S.},
  \bibinfo{author}{Bertrán, J.}, \bibinfo{author}{Moliner, V.},
  \bibinfo{author}{Tuñón, I.}, \bibinfo{year}{2009}.
\newblock \bibinfo{title}{Theoretical modeling of the reaction mechanism of
  phosphate monoester hydrolysis in alkaline phosphatase}.
\newblock \bibinfo{journal}{J. Phys. Chem. B} \bibinfo{volume}{113},
  \bibinfo{pages}{7816--7824}.
\newblock \URLprefix \url{http://dx.doi.org/10.1021/jp901444g},
  \DOIprefix\doi{10.1021/jp901444g},
  \href{http://arxiv.org/abs/http://dx.doi.org/10.1021/jp901444g}{{\tt
  arXiv:http://dx.doi.org/10.1021/jp901444g}}.
\bibitem[{Markland and Manolopoulos(2008a)}]{mark-mano08cpl}
\bibinfo{author}{Markland, T.E.}, \bibinfo{author}{Manolopoulos, D.E.},
  \bibinfo{year}{2008}a.
\newblock \bibinfo{title}{{A refined ring polymer contraction scheme for
  systems with electrostatic interactions}}.
\newblock \bibinfo{journal}{Chem. Phys. Lett.} \bibinfo{volume}{464},
  \bibinfo{pages}{256}.
\bibitem[{Markland and Manolopoulos(2008b)}]{mark-mano08jcp}
\bibinfo{author}{Markland, T.E.}, \bibinfo{author}{Manolopoulos, D.E.},
  \bibinfo{year}{2008}b.
\newblock \bibinfo{title}{{An efficient ring polymer contraction scheme for
  imaginary time path integral simulations.}}
\newblock \bibinfo{journal}{J. Chem. Phys.} \bibinfo{volume}{129},
  \bibinfo{pages}{024105}.
\bibitem[{Marsalek et~al.(2014)Marsalek, Chen, Dupuis, Benoit, Meheut, Bacic
  and Tuckerman}]{Marsalek2014}
\bibinfo{author}{Marsalek, O.}, \bibinfo{author}{Chen, P.Y.},
  \bibinfo{author}{Dupuis, R.}, \bibinfo{author}{Benoit, M.},
  \bibinfo{author}{Meheut, M.}, \bibinfo{author}{Bacic, Z.},
  \bibinfo{author}{Tuckerman, M.E.}, \bibinfo{year}{2014}.
\newblock \bibinfo{title}{Efficient calculation of free energy differences
  associated with isotopic substitution using path-integral molecular
  dynamics}.
\newblock \bibinfo{journal}{J. Chem. Theory Comput.} \bibinfo{volume}{10},
  \bibinfo{pages}{1440--1453}.
\newblock \URLprefix \url{http://dx.doi.org/10.1021/ct400911m},
  \DOIprefix\doi{10.1021/ct400911m}.
\bibitem[{Marsalek and Markland(2016)}]{Marsalek2016}
\bibinfo{author}{Marsalek, O.}, \bibinfo{author}{Markland, T.E.},
  \bibinfo{year}{2016}.
\newblock \bibinfo{title}{Ab initio molecular dynamics with nuclear quantum
  effects at classical cost: Ring polymer contraction for density functional
  theory}.
\newblock \bibinfo{journal}{J. Chem. Phys.} \bibinfo{volume}{144},
  \bibinfo{pages}{054112}.
\bibitem[{Marx and Parrinello(1996)}]{marx-parr95jcp}
\bibinfo{author}{Marx, D.}, \bibinfo{author}{Parrinello, M.},
  \bibinfo{year}{1996}.
\newblock \bibinfo{title}{{Ab initio path integral molecular dynamics: Basic
  ideas}}.
\newblock \bibinfo{journal}{J. Chem. Phys.} \bibinfo{volume}{104},
  \bibinfo{pages}{4077}.
\bibitem[{McKenzie(2012)}]{McKenzie2012}
\bibinfo{author}{McKenzie, R.H.}, \bibinfo{year}{2012}.
\newblock \bibinfo{title}{{A diabatic state model for donor-hydrogen
  vibrational frequency shifts in hydrogen bonded complexes}}.
\newblock \bibinfo{journal}{Chem. Phys. Lett.} \bibinfo{volume}{535},
  \bibinfo{pages}{196--200}.
\bibitem[{Mildvan et~al.(2002)Mildvan, Massiah, Harris, Marks, Harrison,
  Viragh, Reddy and Kovach}]{Mildvan2002}
\bibinfo{author}{Mildvan, A.}, \bibinfo{author}{Massiah, M.},
  \bibinfo{author}{Harris, T.}, \bibinfo{author}{Marks, G.},
  \bibinfo{author}{Harrison, D.}, \bibinfo{author}{Viragh, C.},
  \bibinfo{author}{Reddy, P.}, \bibinfo{author}{Kovach, I.},
  \bibinfo{year}{2002}.
\newblock \bibinfo{title}{Short, strong hydrogen bonds on enzymes: \{NMR\} and
  mechanistic studies}.
\newblock \bibinfo{journal}{J. Mol. Struct.} \bibinfo{volume}{615},
  \bibinfo{pages}{163 -- 175}.
\newblock \URLprefix
  \url{http://www.sciencedirect.com/science/article/pii/S0022286002002120},
  \DOIprefix\doi{http://dx.doi.org/10.1016/S0022-2860(02)00212-0}.
\bibitem[{Mori-Sanchez et~al.({2006})Mori-Sanchez, Cohen and
  Yang}]{Weitao-2006}
\bibinfo{author}{Mori-Sanchez, P.}, \bibinfo{author}{Cohen, A.J.},
  \bibinfo{author}{Yang, W.}, \bibinfo{year}{{2006}}.
\newblock \bibinfo{title}{{Many-electron self-interaction error in approximate
  density functionals}}.
\newblock \bibinfo{journal}{{J. Chem. Phys.}} \bibinfo{volume}{{125}},
  \bibinfo{pages}{{201102}}.
\bibitem[{Mori-S\'{a}nchez et~al.({2008})Mori-S\'{a}nchez, Cohen and
  Yang}]{Weitao-2008}
\bibinfo{author}{Mori-S\'{a}nchez, P.}, \bibinfo{author}{Cohen, A.J.},
  \bibinfo{author}{Yang, W.}, \bibinfo{year}{{2008}}.
\newblock \bibinfo{title}{{Localization and delocalization errors in density
  functional theory and implications for band-gap prediction}}.
\newblock \bibinfo{journal}{{Phys. Rev. Lett.}} \bibinfo{volume}{{100}},
  \bibinfo{pages}{{146401}}.
\bibitem[{Morrone and Car(2008)}]{morr-car08prl}
\bibinfo{author}{Morrone, J.A.}, \bibinfo{author}{Car, R.},
  \bibinfo{year}{2008}.
\newblock \bibinfo{title}{{Nuclear Quantum Effects in Water}}.
\newblock \bibinfo{journal}{Phys. Rev. Lett.} \bibinfo{volume}{101},
  \bibinfo{pages}{17801}.
\bibitem[{Nielsen and McCammon(2003)}]{Nielsen2003}
\bibinfo{author}{Nielsen, J.E.}, \bibinfo{author}{McCammon, J.A.},
  \bibinfo{year}{2003}.
\newblock \bibinfo{title}{Calculating pka values in enzyme active sites}.
\newblock \bibinfo{journal}{Protein Science : A Publication of the Protein
  Society} \bibinfo{volume}{12}, \bibinfo{pages}{1894--1901}.
\newblock \URLprefix
  \url{http://www.ncbi.nlm.nih.gov/pmc/articles/PMC2323987/}.
\bibitem[{Ochterski et~al.(1996)Ochterski, Petersson and
  Montgomery}]{Ochterski_1996}
\bibinfo{author}{Ochterski, J.W.}, \bibinfo{author}{Petersson, G.A.},
  \bibinfo{author}{Montgomery, J.A.}, \bibinfo{year}{1996}.
\newblock \bibinfo{title}{A complete basis set model chemistry. v. extensions
  to six or more heavy atoms}.
\newblock \bibinfo{journal}{J. Chem. Phys.} \bibinfo{volume}{104},
  \bibinfo{pages}{2598--2619}.
\newblock \URLprefix
  \url{http://scitation.aip.org/content/aip/journal/jcp/104/7/10.1063/1.470985},
  \DOIprefix\doi{http://dx.doi.org/10.1063/1.470985}.
\bibitem[{Ordej\'on et~al.(1996)Ordej\'on, Artacho and Soler}]{siesta}
\bibinfo{author}{Ordej\'on, P.}, \bibinfo{author}{Artacho, E.},
  \bibinfo{author}{Soler, J.M.}, \bibinfo{year}{1996}.
\newblock \bibinfo{title}{Self-consistent order-$n$ density-functional
  calculations for very large systems}.
\newblock \bibinfo{journal}{Phys. Rev. B} \bibinfo{volume}{53},
  \bibinfo{pages}{R10441--R10444}.
\newblock \URLprefix \url{http://link.aps.org/doi/10.1103/PhysRevB.53.R10441},
  \DOIprefix\doi{10.1103/PhysRevB.53.R10441}.
\bibitem[{Perdew et~al.(1982)Perdew, Parr, Levy and Balduz}]{perdew1982}
\bibinfo{author}{Perdew, J.P.}, \bibinfo{author}{Parr, R.G.},
  \bibinfo{author}{Levy, M.}, \bibinfo{author}{Balduz, J.L.},
  \bibinfo{year}{1982}.
\newblock \bibinfo{title}{Density-functional theory for fractional particle
  number - derivative discontinuities of the energy}.
\newblock \bibinfo{journal}{Phys. Rev. Lett.} \bibinfo{volume}{49},
  \bibinfo{pages}{1691--1694}.
\bibitem[{Perdew et~al.(2009)Perdew, Ruzsinszky, Constantin, Sun and
  Csonka}]{Perdew_2009}
\bibinfo{author}{Perdew, J.P.}, \bibinfo{author}{Ruzsinszky, A.},
  \bibinfo{author}{Constantin, L.A.}, \bibinfo{author}{Sun, J.},
  \bibinfo{author}{Csonka, G.I.}, \bibinfo{year}{2009}.
\newblock \bibinfo{title}{Some fundamental issues in ground-state density
  functional theory: A guide for the perplexed}.
\newblock \bibinfo{journal}{J. Chem. Theory Comput.} \bibinfo{volume}{5},
  \bibinfo{pages}{902--908}.
\newblock \URLprefix \url{http://dx.doi.org/10.1021/ct800531s},
  \DOIprefix\doi{10.1021/ct800531s},
  \href{http://arxiv.org/abs/http://dx.doi.org/10.1021/ct800531s}{{\tt
  arXiv:http://dx.doi.org/10.1021/ct800531s}}.
\bibitem[{Perdew and Yue(1986)}]{GGA_1986}
\bibinfo{author}{Perdew, J.P.}, \bibinfo{author}{Yue, W.},
  \bibinfo{year}{1986}.
\newblock \bibinfo{title}{Accurate and simple density functional for the
  electronic exchange energy: Generalized gradient approximation}.
\newblock \bibinfo{journal}{Phys. Rev. B} \bibinfo{volume}{33},
  \bibinfo{pages}{8800--8802}.
\newblock \URLprefix \url{http://link.aps.org/doi/10.1103/PhysRevB.33.8800},
  \DOIprefix\doi{10.1103/PhysRevB.33.8800}.
\bibitem[{Perrin and Nielson(1997)}]{Perrin1997}
\bibinfo{author}{Perrin, C.L.}, \bibinfo{author}{Nielson, J.B.},
  \bibinfo{year}{1997}.
\newblock \bibinfo{title}{"strong" hydrogen bonds in chemistry and biology}.
\newblock \bibinfo{journal}{Annual Review of Physical Chemistry}
  \bibinfo{volume}{48}, \bibinfo{pages}{511--544}.
\newblock \URLprefix \url{http://dx.doi.org/10.1146/annurev.physchem.48.1.511},
  \DOIprefix\doi{10.1146/annurev.physchem.48.1.511},
  \href{http://arxiv.org/abs/http://dx.doi.org/10.1146/annurev.physchem.48.1.511}{{\tt
  arXiv:http://dx.doi.org/10.1146/annurev.physchem.48.1.511}}.
\bibitem[{Plimpton(1995)}]{plim95jcp}
\bibinfo{author}{Plimpton, S.}, \bibinfo{year}{1995}.
\newblock \bibinfo{title}{{Fast Parallel Algorithms for Short-Range Molecular
  Dynamics}}.
\newblock \bibinfo{journal}{J. Comp. Phys.} \bibinfo{volume}{117},
  \bibinfo{pages}{1--19}.
\bibitem[{Ponder and Case(2003)}]{Ponder2003}
\bibinfo{author}{Ponder, J.W.}, \bibinfo{author}{Case, D.A.},
  \bibinfo{year}{2003}.
\newblock \bibinfo{title}{Force fields for protein simulations}.
\newblock \bibinfo{journal}{Advances in Protein Chemistry}
  \bibinfo{volume}{66}, \bibinfo{pages}{27 -- 85}.
\bibitem[{Ranaghan and Mulholland(2010)}]{Ranaghan_2010}
\bibinfo{author}{Ranaghan, K.E.}, \bibinfo{author}{Mulholland, A.J.},
  \bibinfo{year}{2010}.
\newblock \bibinfo{title}{Investigations of enzyme-catalysed reactions with
  combined quantum mechanics/molecular mechanics (qm/mm) methods}.
\newblock \bibinfo{journal}{International Reviews in Physical Chemistry}
  \bibinfo{volume}{29}, \bibinfo{pages}{65--133}.
\newblock \URLprefix \url{http://dx.doi.org/10.1080/01442350903495417},
  \DOIprefix\doi{10.1080/01442350903495417},
  \href{http://arxiv.org/abs/http://dx.doi.org/10.1080/01442350903495417}{{\tt
  arXiv:http://dx.doi.org/10.1080/01442350903495417}}.
\bibitem[{Refaely-Abramson et~al.({2011})Refaely-Abramson, Baer and
  Kronik}]{Baer-2011}
\bibinfo{author}{Refaely-Abramson, S.}, \bibinfo{author}{Baer, R.},
  \bibinfo{author}{Kronik, L.}, \bibinfo{year}{{2011}}.
\newblock \bibinfo{title}{{Fundamental and excitation gaps in molecules of
  relevance for organic photovoltaics from an optimally tuned range-separated
  hybrid functional}}.
\newblock \bibinfo{journal}{{Phys. Rev. B}} \bibinfo{volume}{{84}},
  \bibinfo{pages}{{075144}}.
\bibitem[{Refaely-Abramson et~al.({2012})Refaely-Abramson, Sharifzadeh, Govind,
  Autschbach, Neaton, Baer and Kronik}]{Baer-2012-b}
\bibinfo{author}{Refaely-Abramson, S.}, \bibinfo{author}{Sharifzadeh, S.},
  \bibinfo{author}{Govind, N.}, \bibinfo{author}{Autschbach, J.},
  \bibinfo{author}{Neaton, J.B.}, \bibinfo{author}{Baer, R.},
  \bibinfo{author}{Kronik, L.}, \bibinfo{year}{{2012}}.
\newblock \bibinfo{title}{{Quasiparticle Spectra from a Nonempirical Optimally
  Tuned Range-Separated Hybrid Density Functional}}.
\newblock \bibinfo{journal}{{Phys. Rev. Lett.}} \bibinfo{volume}{{109}},
  \bibinfo{pages}{{226405}}.
\bibitem[{Riccardi et~al.(2006)Riccardi, Schaefer, Yang, Yu, Ghosh,
  Prat-Resina, König, Li, Xu, Guo, Elstner and Cui}]{Riccardi_2006}
\bibinfo{author}{Riccardi, D.}, \bibinfo{author}{Schaefer, P.},
  \bibinfo{author}{Yang, Y.}, \bibinfo{author}{Yu, H.}, \bibinfo{author}{Ghosh,
  N.}, \bibinfo{author}{Prat-Resina, X.}, \bibinfo{author}{König, P.},
  \bibinfo{author}{Li, G.}, \bibinfo{author}{Xu, D.}, \bibinfo{author}{Guo,
  H.}, \bibinfo{author}{Elstner, M.}, \bibinfo{author}{Cui, Q.},
  \bibinfo{year}{2006}.
\newblock \bibinfo{title}{Development of effective quantum mechanical/molecular
  mechanical (qm/mm) methods for complex biological processes}.
\newblock \bibinfo{journal}{J. Phys. Chem. B} \bibinfo{volume}{110},
  \bibinfo{pages}{6458--6469}.
\newblock \URLprefix \url{http://dx.doi.org/10.1021/jp056361o},
  \DOIprefix\doi{10.1021/jp056361o},
  \href{http://arxiv.org/abs/http://dx.doi.org/10.1021/jp056361o}{{\tt
  arXiv:http://dx.doi.org/10.1021/jp056361o}}.
\bibitem[{Salzner and Baer({2009})}]{Baer-2009-c}
\bibinfo{author}{Salzner, U.}, \bibinfo{author}{Baer, R.},
  \bibinfo{year}{{2009}}.
\newblock \bibinfo{title}{{Koopmans' springs to life}}.
\newblock \bibinfo{journal}{{J. Chem. Phys.}} \bibinfo{volume}{{131}},
  \bibinfo{pages}{{231101}}.
\bibitem[{Schmidt et~al.({2014})Schmidt, Kraisler, Makmal, Kronik and
  Kuemmel}]{Kronik-2014-b}
\bibinfo{author}{Schmidt, T.}, \bibinfo{author}{Kraisler, E.},
  \bibinfo{author}{Makmal, A.}, \bibinfo{author}{Kronik, L.},
  \bibinfo{author}{Kuemmel, S.}, \bibinfo{year}{{2014}}.
\newblock \bibinfo{title}{{A self-interaction-free local hybrid functional:
  Accurate binding energies vis-a-vis accurate ionization potentials from
  Kohn-Sham eigenvalues}}.
\newblock \bibinfo{journal}{{J. Chem. Phys.}} \bibinfo{volume}{{140}},
  \bibinfo{pages}{{18A510}}.
\bibitem[{Seidl et~al.(1996)Seidl, G\"orling, Vogl, Majewski and
  Levy}]{Seidl_1996}
\bibinfo{author}{Seidl, A.}, \bibinfo{author}{G\"orling, A.},
  \bibinfo{author}{Vogl, P.}, \bibinfo{author}{Majewski, J.A.},
  \bibinfo{author}{Levy, M.}, \bibinfo{year}{1996}.
\newblock \bibinfo{title}{Generalized kohn-sham schemes and the band-gap
  problem}.
\newblock \bibinfo{journal}{Phys. Rev. B} \bibinfo{volume}{53},
  \bibinfo{pages}{3764--3774}.
\newblock \URLprefix \url{http://link.aps.org/doi/10.1103/PhysRevB.53.3764},
  \DOIprefix\doi{10.1103/PhysRevB.53.3764}.
\bibitem[{Senn and Thiel(2009)}]{Hans_Martin_2009}
\bibinfo{author}{Senn, H.M.}, \bibinfo{author}{Thiel, W.},
  \bibinfo{year}{2009}.
\newblock \bibinfo{title}{Qm/mm methods for biomolecular systems}.
\newblock \bibinfo{journal}{Angewandte Chemie International Edition}
  \bibinfo{volume}{48}, \bibinfo{pages}{1198--1229}.
\newblock \URLprefix \url{http://dx.doi.org/10.1002/anie.200802019},
  \DOIprefix\doi{10.1002/anie.200802019}.
\bibitem[{Siegbahn and Borowski(2006)}]{Siegbahn2006}
\bibinfo{author}{Siegbahn, P.E.M.}, \bibinfo{author}{Borowski, T.},
  \bibinfo{year}{2006}.
\newblock \bibinfo{title}{Modeling enzymatic reactions involving transition
  metals}.
\newblock \bibinfo{journal}{Accounts of Chemical Research}
  \bibinfo{volume}{39}, \bibinfo{pages}{729--738}.
\newblock \URLprefix \url{http://dx.doi.org/10.1021/ar050123u},
  \DOIprefix\doi{10.1021/ar050123u},
  \href{http://arxiv.org/abs/http://dx.doi.org/10.1021/ar050123u}{{\tt
  arXiv:http://dx.doi.org/10.1021/ar050123u}}.
\bibitem[{Sigala et~al.(2013)Sigala, Fafarman, Schwans, Fried, Fenn, Caaveiro,
  Pybus, Ringe, Petsko, Boxer and Herschlag}]{Sigala2013}
\bibinfo{author}{Sigala, P.A.}, \bibinfo{author}{Fafarman, A.T.},
  \bibinfo{author}{Schwans, J.P.}, \bibinfo{author}{Fried, S.D.},
  \bibinfo{author}{Fenn, T.D.}, \bibinfo{author}{Caaveiro, J.M.M.},
  \bibinfo{author}{Pybus, B.}, \bibinfo{author}{Ringe, D.},
  \bibinfo{author}{Petsko, G.A.}, \bibinfo{author}{Boxer, S.G.},
  \bibinfo{author}{Herschlag, D.}, \bibinfo{year}{2013}.
\newblock \bibinfo{title}{Quantitative dissection of hydrogen bond-mediated
  proton transfer in the ketosteroid isomerase active site}.
\newblock \bibinfo{journal}{Proc. Natl. Acad. Sci. U.S.A.}
  \bibinfo{volume}{110}, \bibinfo{pages}{E2552--E2561}.
\newblock \URLprefix \url{http://www.pnas.org/content/110/28/E2552.abstract}.
\bibitem[{Simonson et~al.(2002)Simonson, Archontis and Karplus}]{Simonson2002}
\bibinfo{author}{Simonson, T.}, \bibinfo{author}{Archontis, G.},
  \bibinfo{author}{Karplus, M.}, \bibinfo{year}{2002}.
\newblock \bibinfo{title}{Free energy simulations come of age: Protein-ligand
  recognition}.
\newblock \bibinfo{journal}{Accounts of Chemical Research}
  \bibinfo{volume}{35}, \bibinfo{pages}{430--437}.
\newblock \URLprefix \url{http://dx.doi.org/10.1021/ar010030m},
  \DOIprefix\doi{10.1021/ar010030m},
  \href{http://arxiv.org/abs/http://dx.doi.org/10.1021/ar010030m}{{\tt
  arXiv:http://dx.doi.org/10.1021/ar010030m}}.
\bibitem[{Skylaris et~al.(2005)Skylaris, Haynes, Mostofi and Payne}]{onetep}
\bibinfo{author}{Skylaris, C.K.}, \bibinfo{author}{Haynes, P.D.},
  \bibinfo{author}{Mostofi, A.A.}, \bibinfo{author}{Payne, M.C.},
  \bibinfo{year}{2005}.
\newblock \bibinfo{title}{Introducing onetep: Linear-scaling density functional
  simulations on parallel computers}.
\newblock \bibinfo{journal}{J. Chem. Phys.} \bibinfo{volume}{122}.
\newblock \URLprefix
  \url{http://scitation.aip.org/content/aip/journal/jcp/122/8/10.1063/1.1839852},
  \DOIprefix\doi{http://dx.doi.org/10.1063/1.1839852}.
\bibitem[{Sosa~Vazquez and Isborn(2015)}]{Sosa_Vazquez_2015}
\bibinfo{author}{Sosa~Vazquez, X.A.}, \bibinfo{author}{Isborn, C.M.},
  \bibinfo{year}{2015}.
\newblock \bibinfo{title}{Size-dependent error of the density functional theory
  ionization potential in vacuum and solution}.
\newblock \bibinfo{journal}{J. Chem. Phys.} \bibinfo{volume}{143}.
\newblock \URLprefix
  \url{http://scitation.aip.org/content/aip/journal/jcp/143/24/10.1063/1.4937417},
  \DOIprefix\doi{http://dx.doi.org/10.1063/1.4937417}.
\bibitem[{Stein et~al.({2012})Stein, Autschbach, Govind, Kronik and
  Baer}]{Baer-2012}
\bibinfo{author}{Stein, T.}, \bibinfo{author}{Autschbach, J.},
  \bibinfo{author}{Govind, N.}, \bibinfo{author}{Kronik, L.},
  \bibinfo{author}{Baer, R.}, \bibinfo{year}{{2012}}.
\newblock \bibinfo{title}{{Curvature and Frontier Orbital Energies in Density
  Functional Theory}}.
\newblock \bibinfo{journal}{{J. Phys. Chem. Lett.}} \bibinfo{volume}{{3}},
  \bibinfo{pages}{{3740--3744}}.
\bibitem[{Stein et~al.({2010})Stein, Eisenberg, Kronik and Baer}]{Baer-2010}
\bibinfo{author}{Stein, T.}, \bibinfo{author}{Eisenberg, H.},
  \bibinfo{author}{Kronik, L.}, \bibinfo{author}{Baer, R.},
  \bibinfo{year}{{2010}}.
\newblock \bibinfo{title}{{Fundamental Gaps in Finite Systems from Eigenvalues
  of a Generalized Kohn-Sham Method}}.
\newblock \bibinfo{journal}{{Phys. Rev. Lett.}} \bibinfo{volume}{{105}},
  \bibinfo{pages}{{266802}}.
\bibitem[{Sumowski and Ochsenfeld(2009)}]{Sumowski_2009}
\bibinfo{author}{Sumowski, C.V.}, \bibinfo{author}{Ochsenfeld, C.},
  \bibinfo{year}{2009}.
\newblock \bibinfo{title}{A convergence study of qm/mm isomerization energies
  with the selected size of the qm region for peptidic systems}.
\newblock \bibinfo{journal}{J. Phys. Chem. A} \bibinfo{volume}{113},
  \bibinfo{pages}{11734--11741}.
\newblock \URLprefix \url{http://dx.doi.org/10.1021/jp902876n},
  \DOIprefix\doi{10.1021/jp902876n},
  \href{http://arxiv.org/abs/http://dx.doi.org/10.1021/jp902876n}{{\tt
  arXiv:http://dx.doi.org/10.1021/jp902876n}}.
\bibitem[{Sutcliffe and Scrutton(2002)}]{Sutcliffe2002}
\bibinfo{author}{Sutcliffe, M.J.}, \bibinfo{author}{Scrutton, N.S.},
  \bibinfo{year}{2002}.
\newblock \bibinfo{title}{A new conceptual framework for enzyme catalysis}.
\newblock \bibinfo{journal}{Eur. J. Biochem.} \bibinfo{volume}{269},
  \bibinfo{pages}{3096--3102}.
\newblock \URLprefix \url{http://dx.doi.org/10.1046/j.1432-1033.2002.03020.x}.
\bibitem[{Tao et~al.(2003)Tao, Perdew, Staroverov and Scuseria}]{TPSS_2003}
\bibinfo{author}{Tao, J.}, \bibinfo{author}{Perdew, J.P.},
  \bibinfo{author}{Staroverov, V.N.}, \bibinfo{author}{Scuseria, G.E.},
  \bibinfo{year}{2003}.
\newblock \bibinfo{title}{Climbing the density functional ladder: Nonempirical
  meta\char21{}generalized gradient approximation designed for molecules and
  solids}.
\newblock \bibinfo{journal}{Phys. Rev. Lett.} \bibinfo{volume}{91},
  \bibinfo{pages}{146401}.
\newblock \URLprefix
  \url{http://link.aps.org/doi/10.1103/PhysRevLett.91.146401},
  \DOIprefix\doi{10.1103/PhysRevLett.91.146401}.
\bibitem[{Tawada et~al.(2004)Tawada, Tsuneda, Yanagisawa, Yanai and
  Hirao}]{Hirao-2004}
\bibinfo{author}{Tawada, Y.}, \bibinfo{author}{Tsuneda, T.},
  \bibinfo{author}{Yanagisawa, S.}, \bibinfo{author}{Yanai, T.},
  \bibinfo{author}{Hirao, K.}, \bibinfo{year}{2004}.
\newblock \bibinfo{title}{A long-range-corrected time-dependent density
  functional theory}.
\newblock \bibinfo{journal}{J. Chem. Phys.} \bibinfo{volume}{120},
  \bibinfo{pages}{8425--8433}.
\bibitem[{Toulouse et~al.(2004)Toulouse, Colonna and Savin}]{Savin-2004}
\bibinfo{author}{Toulouse, J.}, \bibinfo{author}{Colonna, F.},
  \bibinfo{author}{Savin, A.}, \bibinfo{year}{2004}.
\newblock \bibinfo{title}{Long-range\char21{}short-range separation of the
  electron-electron interaction in density-functional theory}.
\newblock \bibinfo{journal}{Phys. Rev. A} \bibinfo{volume}{70},
  \bibinfo{pages}{062505}.
\bibitem[{Tozer(2003)}]{Tozer_2003}
\bibinfo{author}{Tozer, D.J.}, \bibinfo{year}{2003}.
\newblock \bibinfo{title}{Relationship between long-range charge-transfer
  excitation energy error and integer discontinuity in kohn-sham theory}.
\newblock \bibinfo{journal}{J. Chem. Phys.} \bibinfo{volume}{119},
  \bibinfo{pages}{12697--12699}.
\newblock \URLprefix
  \url{http://scitation.aip.org/content/aip/journal/jcp/119/24/10.1063/1.1633756},
  \DOIprefix\doi{http://dx.doi.org/10.1063/1.1633756}.
\bibitem[{Ufimtsev et~al.(2011)Ufimtsev, Luehr and Martinez}]{Ufimtsev_2011}
\bibinfo{author}{Ufimtsev, I.S.}, \bibinfo{author}{Luehr, N.},
  \bibinfo{author}{Martinez, T.J.}, \bibinfo{year}{2011}.
\newblock \bibinfo{title}{Charge transfer and polarization in solvated proteins
  from ab initio molecular dynamics}.
\newblock \bibinfo{journal}{J. Phys. Chem. Lett.} \bibinfo{volume}{2},
  \bibinfo{pages}{1789--1793}.
\newblock \URLprefix \url{http://dx.doi.org/10.1021/jz200697c},
  \DOIprefix\doi{10.1021/jz200697c},
  \href{http://arxiv.org/abs/http://dx.doi.org/10.1021/jz200697c}{{\tt
  arXiv:http://dx.doi.org/10.1021/jz200697c}}.
\bibitem[{Ufimtsev and Martinez(2008)}]{Ufimtsev_2008}
\bibinfo{author}{Ufimtsev, I.S.}, \bibinfo{author}{Martinez, T.J.},
  \bibinfo{year}{2008}.
\newblock \bibinfo{title}{Quantum chemistry on graphical processing units. 1.
  strategies for two-electron integral evaluation}.
\newblock \bibinfo{journal}{J. Chem. Theory Comput.} \bibinfo{volume}{4},
  \bibinfo{pages}{222--231}.
\newblock \URLprefix \url{http://dx.doi.org/10.1021/ct700268q},
  \DOIprefix\doi{10.1021/ct700268q},
  \href{http://arxiv.org/abs/http://dx.doi.org/10.1021/ct700268q}{{\tt
  arXiv:http://dx.doi.org/10.1021/ct700268q}}.
\bibitem[{Ufimtsev and Martinez(2009a)}]{Ufimtsev_2009a}
\bibinfo{author}{Ufimtsev, I.S.}, \bibinfo{author}{Martinez, T.J.},
  \bibinfo{year}{2009}a.
\newblock \bibinfo{title}{Quantum chemistry on graphical processing units. 2.
  direct self-consistent-field implementation}.
\newblock \bibinfo{journal}{J. Chem. Theory Comput.} \bibinfo{volume}{5},
  \bibinfo{pages}{1004--1015}.
\newblock \URLprefix \url{http://dx.doi.org/10.1021/ct800526s},
  \DOIprefix\doi{10.1021/ct800526s},
  \href{http://arxiv.org/abs/http://dx.doi.org/10.1021/ct800526s}{{\tt
  arXiv:http://dx.doi.org/10.1021/ct800526s}}.
\bibitem[{Ufimtsev and Martinez(2009b)}]{Ufimtsev_2009b}
\bibinfo{author}{Ufimtsev, I.S.}, \bibinfo{author}{Martinez, T.J.},
  \bibinfo{year}{2009}b.
\newblock \bibinfo{title}{Quantum chemistry on graphical processing units. 3.
  analytical energy gradients, geometry optimization, and first principles
  molecular dynamics}.
\newblock \bibinfo{journal}{J. Chem. Theory Comput.} \bibinfo{volume}{5},
  \bibinfo{pages}{2619--2628}.
\newblock \URLprefix \url{http://dx.doi.org/10.1021/ct9003004},
  \DOIprefix\doi{10.1021/ct9003004},
  \href{http://arxiv.org/abs/http://dx.doi.org/10.1021/ct9003004}{{\tt
  arXiv:http://dx.doi.org/10.1021/ct9003004}}.
\bibitem[{Vanicek and Miller(2007)}]{Vanicek2007}
\bibinfo{author}{Vanicek, J.}, \bibinfo{author}{Miller, W.H.},
  \bibinfo{year}{2007}.
\newblock \bibinfo{title}{Efficient estimators for quantum instanton evaluation
  of the kinetic isotope effects: Application to the intramolecular hydrogen
  transfer in pentadiene}.
\newblock \bibinfo{journal}{J. Chem. Phys.} \bibinfo{volume}{127},
  \bibinfo{pages}{114309}.
\newblock \URLprefix
  \url{http://scitation.aip.org/content/aip/journal/jcp/127/11/10.1063/1.2768930},
  \DOIprefix\doi{http://dx.doi.org/10.1063/1.2768930}.
\bibitem[{Wallqvist and Berne(1985)}]{Wallqvist1985}
\bibinfo{author}{Wallqvist, A.}, \bibinfo{author}{Berne, B.},
  \bibinfo{year}{1985}.
\newblock \bibinfo{title}{Path-integral simulation of pure water}.
\newblock \bibinfo{journal}{Chem. Phys. Lett.} \bibinfo{volume}{117},
  \bibinfo{pages}{214 -- 219}.
\newblock \URLprefix
  \url{http://www.sciencedirect.com/science/article/pii/0009261485802062},
  \DOIprefix\doi{http://dx.doi.org/10.1016/0009-2614(85)80206-2}.
\bibitem[{Wang et~al.(2014a)Wang, Ceriotti and Markland}]{Wang2014}
\bibinfo{author}{Wang, L.}, \bibinfo{author}{Ceriotti, M.},
  \bibinfo{author}{Markland, T.E.}, \bibinfo{year}{2014}a.
\newblock \bibinfo{title}{Quantum fluctuations and isotope effects in ab initio
  descriptions of water}.
\newblock \bibinfo{journal}{J. Chem. Phys.} \bibinfo{volume}{141},
  \bibinfo{pages}{104502}.
\newblock \URLprefix
  \url{http://scitation.aip.org/content/aip/journal/jcp/141/10/10.1063/1.4894287},
  \DOIprefix\doi{http://dx.doi.org/10.1063/1.4894287}.
\bibitem[{Wang et~al.(2014b)Wang, Fried, Boxer and Markland}]{Wang2014a}
\bibinfo{author}{Wang, L.}, \bibinfo{author}{Fried, S.D.},
  \bibinfo{author}{Boxer, S.G.}, \bibinfo{author}{Markland, T.E.},
  \bibinfo{year}{2014}b.
\newblock \bibinfo{title}{Quantum delocalization of protons in the
  hydrogen-bond network of an enzyme active site}.
\newblock \bibinfo{journal}{Proc. Natl. Acad. Sci. U.S.A.}
  \bibinfo{volume}{111}, \bibinfo{pages}{18454--18459}.
\newblock \URLprefix \url{http://www.pnas.org/content/111/52/18454}.
\bibitem[{Wang et~al.(2014c)Wang, Titov, McGibbon, Liu, Pande and
  Mart{\'\i}nez}]{Wang_2014}
\bibinfo{author}{Wang, L.P.}, \bibinfo{author}{Titov, A.},
  \bibinfo{author}{McGibbon, R.}, \bibinfo{author}{Liu, F.},
  \bibinfo{author}{Pande, V.S.}, \bibinfo{author}{Mart{\'\i}nez, T.J.},
  \bibinfo{year}{2014}c.
\newblock \bibinfo{title}{Discovering chemistry with an ab initio nanoreactor}.
\newblock \bibinfo{journal}{Nat Chem} \bibinfo{volume}{6},
  \bibinfo{pages}{1044--1048}.
\newblock \URLprefix \url{http://dx.doi.org/10.1038/nchem.2099}.
\bibitem[{Wang and Wilson(2004)}]{Wang2004}
\bibinfo{author}{Wang, N.X.}, \bibinfo{author}{Wilson, A.K.},
  \bibinfo{year}{2004}.
\newblock \bibinfo{title}{The behavior of density functionals with respect to
  basis set. i. the correlation consistent basis sets}.
\newblock \bibinfo{journal}{J. Chem. Phys.} \bibinfo{volume}{121},
  \bibinfo{pages}{7632--7646}.
\newblock \URLprefix
  \url{http://scitation.aip.org/content/aip/journal/jcp/121/16/10.1063/1.1792071},
  \DOIprefix\doi{http://dx.doi.org/10.1063/1.1792071}.
\bibitem[{Warshel and Levitt(1976)}]{Warshel_1976}
\bibinfo{author}{Warshel, A.}, \bibinfo{author}{Levitt, M.},
  \bibinfo{year}{1976}.
\newblock \bibinfo{title}{Theoretical studies of enzymic reactions: Dielectric,
  electrostatic and steric stabilization of the carbonium ion in the reaction
  of lysozyme}.
\newblock \bibinfo{journal}{J. Mol. Biol.} \bibinfo{volume}{103},
  \bibinfo{pages}{227 -- 249}.
\newblock \URLprefix
  \url{http://www.sciencedirect.com/science/article/pii/0022283676903119},
  \DOIprefix\doi{http://dx.doi.org/10.1016/0022-2836(76)90311-9}.
\bibitem[{Whittleton et~al.(2015)Whittleton, Sosa~Vazquez, Isborn and
  Johnson}]{Whittleton_2015}
\bibinfo{author}{Whittleton, S.R.}, \bibinfo{author}{Sosa~Vazquez, X.A.},
  \bibinfo{author}{Isborn, C.M.}, \bibinfo{author}{Johnson, E.R.},
  \bibinfo{year}{2015}.
\newblock \bibinfo{title}{Density-functional errors in ionization potential
  with increasing system size}.
\newblock \bibinfo{journal}{J. Chem. Phys.} \bibinfo{volume}{142},
  \bibinfo{pages}{184106}.
\newblock \URLprefix
  \url{http://scitation.aip.org/content/aip/journal/jcp/142/18/10.1063/1.4920947},
  \DOIprefix\doi{http://dx.doi.org/10.1063/1.4920947}.
\bibitem[{Wodrich et~al.(2008)Wodrich, Jana, von Ragué~Schleyer and
  Corminboeuf}]{Corminboeuf_2008}
\bibinfo{author}{Wodrich, M.D.}, \bibinfo{author}{Jana, D.F.},
  \bibinfo{author}{von Ragué~Schleyer, P.}, \bibinfo{author}{Corminboeuf, C.},
  \bibinfo{year}{2008}.
\newblock \bibinfo{title}{Empirical corrections to density functional theory
  highlight the importance of nonbonded intramolecular interactions in
  alkanes}.
\newblock \bibinfo{journal}{J. Phys. Chem. A} \bibinfo{volume}{112},
  \bibinfo{pages}{11495--11500}.
\newblock \URLprefix \url{http://dx.doi.org/10.1021/jp806619z},
  \DOIprefix\doi{10.1021/jp806619z},
  \href{http://arxiv.org/abs/http://dx.doi.org/10.1021/jp806619z}{{\tt
  arXiv:http://dx.doi.org/10.1021/jp806619z}}.
\bibitem[{Wu et~al.(2001)Wu, Vargas, Nayak, Lotrich and Scoles}]{Wu_2001}
\bibinfo{author}{Wu, X.}, \bibinfo{author}{Vargas, M.C.},
  \bibinfo{author}{Nayak, S.}, \bibinfo{author}{Lotrich, V.},
  \bibinfo{author}{Scoles, G.}, \bibinfo{year}{2001}.
\newblock \bibinfo{title}{Towards extending the applicability of density
  functional theory to weakly bound systems}.
\newblock \bibinfo{journal}{J. Chem. Phys.} \bibinfo{volume}{115},
  \bibinfo{pages}{8748--8757}.
\newblock \URLprefix
  \url{http://scitation.aip.org/content/aip/journal/jcp/115/19/10.1063/1.1412004},
  \DOIprefix\doi{http://dx.doi.org/10.1063/1.1412004}.
\bibitem[{Yanai et~al.({2004})Yanai, Tew and Handy}]{Handy-2004}
\bibinfo{author}{Yanai, T.}, \bibinfo{author}{Tew, D.}, \bibinfo{author}{Handy,
  N.}, \bibinfo{year}{{2004}}.
\newblock \bibinfo{title}{{A new hybrid exchange-correlation functional using
  the Coulomb-attenuating method (CAM-B3LYP)}}.
\newblock \bibinfo{journal}{{Chemical Physics Letters}}
  \bibinfo{volume}{{393}}, \bibinfo{pages}{{51--57}}.
\bibitem[{Zhang et~al.(2015)Zhang, Kulik, Martinez and Klinman}]{Zhang_2015}
\bibinfo{author}{Zhang, J.}, \bibinfo{author}{Kulik, H.J.},
  \bibinfo{author}{Martinez, T.J.}, \bibinfo{author}{Klinman, J.P.},
  \bibinfo{year}{2015}.
\newblock \bibinfo{title}{Mediation of donor–acceptor distance in an
  enzymatic methyl transfer reaction}.
\newblock \bibinfo{journal}{Proc. Natl. Acad. Sci. U.S.A.}
  \bibinfo{volume}{112}, \bibinfo{pages}{7954--7959}.
\bibitem[{Zhang and Yang({1998})}]{Weitao-1998}
\bibinfo{author}{Zhang, Y.}, \bibinfo{author}{Yang, W.},
  \bibinfo{year}{{1998}}.
\newblock \bibinfo{title}{{A challenge for density functionals:
  Self-interaction error increases for systems with a noninteger number of
  electrons}}.
\newblock \bibinfo{journal}{{J. Chem. Phys.}} \bibinfo{volume}{{109}},
  \bibinfo{pages}{{2604--2608}}.
\bibitem[{Zhao et~al.(2015)Zhao, Xie and Kulik}]{Zhao_2015}
\bibinfo{author}{Zhao, Q.}, \bibinfo{author}{Xie, L.}, \bibinfo{author}{Kulik,
  H.J.}, \bibinfo{year}{2015}.
\newblock \bibinfo{title}{Discovering amorphous indium phosphide nanostructures
  with high-temperature ab initio molecular dynamics}.
\newblock \bibinfo{journal}{J. Phys. Chem. C} \bibinfo{volume}{119},
  \bibinfo{pages}{23238--23249}.
\newblock \URLprefix \url{http://dx.doi.org/10.1021/acs.jpcc.5b07264},
  \DOIprefix\doi{10.1021/acs.jpcc.5b07264},
  \href{http://arxiv.org/abs/http://dx.doi.org/10.1021/acs.jpcc.5b07264}{{\tt
  arXiv:http://dx.doi.org/10.1021/acs.jpcc.5b07264}}.
\bibitem[{Zheng et~al.({2012})Zheng, Liu, Johnson, Contreras-Garcia and
  Yang}]{Weitao-2012}
\bibinfo{author}{Zheng, X.}, \bibinfo{author}{Liu, M.},
  \bibinfo{author}{Johnson, E.R.}, \bibinfo{author}{Contreras-Garcia, J.},
  \bibinfo{author}{Yang, W.}, \bibinfo{year}{{2012}}.
\newblock \bibinfo{title}{{Delocalization error of density-functional
  approximations: A distinct manifestation in hydrogen molecular chains}}.
\newblock \bibinfo{journal}{{J. Chem. Phys.}} \bibinfo{volume}{{137}},
  \bibinfo{pages}{{214106}}.

\end{thebibliography}

\end{document}